\documentclass[twocolumn]{aastex63}
\pdfoutput=1

\usepackage{soul}
\usepackage[version=4]{mhchem}

\usepackage[separate-uncertainty=true, multi-part-units=single,  range-units=single, retain-explicit-plus=true]{siunitx}

\let\cite\citep
\let\textcite\citet

\newcommand{\coords}[2]{\ang{#1}{W},~\ang{#2}{N}}

\submitjournal{PSJ}
\shorttitle{Compositional Mapping of Europa with VLT/SPHERE and Galileo/NIMS Observations}
\shortauthors{King et al.}
\begin{document}

\title{Compositional Mapping of Europa using MCMC Modelling of Near-IR VLT/SPHERE and Galileo/NIMS Observations}
\correspondingauthor{Oliver King}
\email{ortk1@le.ac.uk}
\author[0000-0002-6271-0062]{Oliver King}
\affiliation{School of Physics and Astronomy, University of Leicester, University Road, Leicester, LE1 7RH, United Kingdom}
\author[0000-0001-5834-9588]{Leigh N Fletcher}
\affiliation{School of Physics and Astronomy, University of Leicester, University Road, Leicester, LE1 7RH, United Kingdom}
\author[0000-0001-7212-6241]{Nicolas Ligier}
\affiliation{Université Paris-Saclay, CNRS, Institut d’Astrophysique Spatiale, 91405, Orsay, France}

\begin{abstract}
	We present maps of surface composition of Europa's anti-jovian hemisphere acquired using high spatial resolution IFU multi-spectral data from the SPHERE instrument on the Very Large Telescope (\SIrange{0.95}{1.65}{\micro\m}) and the NIMS instrument on the Galileo orbiter (\SIrange{0.7}{5.2}{\micro\m}). Spectral modelling was performed using a Markov Chain Monte Carlo method to estimate endmember abundances and to quantify their associated uncertainties. Modelling results support the leading-trailing hemisphere difference in hydrated sulphuric acid abundances caused by exogenic plasma bombardment. Water ice grains are found to be in the \SI{100}{\micro\m} to \SI{1}{mm} range, with larger grains present on the trailing hemisphere, consistent with radiation driven sputtering destroying smaller grains. Modelling best estimates suggest a mixture of sulphate and chlorinated salts, although uncertainties derived from the MCMC modelling suggest that it is difficult to confidently detect individual salt abundances with low spectral resolution spectra from SPHERE and NIMS. The high spatial resolution offered by SPHERE allows the small scale spatial distribution (\SI{<150}{km}) of potential species to be mapped, including ground-based detection of lineae and impact features. This could be used in combination with other higher spectral resolution observations to confirm the presence of these species.
\end{abstract}


\section{Introduction}
\label{sec:intro}
The Galilean moons - Io, Europa, Ganymede and Callisto - are Jupiter's four largest moons. The moons collectively form a solar system in miniature around Jupiter, with environments ranging from the volcanic and rocky Io to the icy Europa, Ganymede and Callisto.

Europa, the second Galilean moon from Jupiter, is the smallest of the four moons. It has a core of silicate rock with an outer crust of liquid water and water ice which may only be \SI{\sim20}{km} thick \cite{howell2021likely}. Europa is tidally locked, meaning it always presents the same face to Jupiter, and tidal heating is sufficient to maintain a liquid subsurface ocean between the surface and silicate interior \cite{greeley2004geology, schubert2004interior}. The presence of a liquid (i.e. current-carrying) subsurface ocean is supported by induced magnetic field measurements by Galileo \cite{kivelson2000galileo}. The subsurface ocean and its direct contact with Europa's silicate interior makes Europa one of the most likely candidates in the solar system to be able to support habitable conditions \cite{chyba2000energy, marion2003search}.

Europa is geologically active, with transient cryovolcanic plumes of water vapour tentatively detected in Hubble Space Telescope observations \cite{roth2014transient}. It has a very smooth surface, with few impact craters, implying that the surface is geologically very young, with an average age of \SI{\sim50}{Myr}. This young surface age implies that recent cryovolcanic resurfacing events could leave detectable signatures from the subsurface ocean on Europa's surface, potentially allowing infrared spectroscopy to identify species that may be present in the ocean, albeit chemically processed via exposure on the surface \cite{pappalardo1999does, greeley2004geology}.

Europa's surface is mainly composed of water ice, and is covered with a series of intersecting linear features or `lineae', the largest of which are over \SI{1000}{km} long. These lineae are thought to be caused by tidal stresses on Europa's surface which opens fissures, exposing the warmer ice layers beneath \cite{greeley2004geology}. `Chaos' terrains are composed of irregular polygonal blocks of old surface material that are set in younger ice. These are also thought to be caused by endogenic processes, with some form of melting or softening of the ice crust, though the exact mechanism is not fully understood \cite{greeley2004geology, figueredo2004resurfacing}.

Much of the current understanding of Europa's surface composition comes from infrared spectroscopic observations. Observed reflectance spectra can be compared to reference spectra measured in laboratories on Earth to identify the cause of distinctive absorption features in Europa's spectrum. The most comprehensive study of Europa comes from the Galileo orbiter mission that orbited Jupiter from 1995 to 2003 with repeated flybys of the Galilean satellites. Galileo's Near-Infrared Mapping Spectrometer, NIMS, (\SIrange{0.7}{5.2}{\micro\m}, $R = \lambda/\Delta\lambda \sim 60$) \cite{carlson1992near} confirmed a surface dominated by water ice although with significant amounts non-ice contamination. Many other spacecraft have taken observations during flybys of the Jovian system, including the New Horizons flyby in 2007 which allowed observations of Europa with the LEISA \cite{reuter2008ralph} instrument (\SIrange{1.25}{2.5}{\micro\m}, $R=240$) \cite{grundy2007new}.

One of the key areas of study is the composition of the non-ice material on Europa's surface causing distortions to the detected water-ice absorption bands. The largest component of the non-ice material is hydrated sulphuric acid, which can account for much of the contamination on Europa's trailing hemisphere. Observations by NIMS show a strong correlation between the sulphuric acid distribution and the `bullseye' distribution jovian plasma bombardment which is centred on Europa's trailing apex (\coords{270}{0}) suggesting an exogenic origin for the sulphuric acid \cite{carlson2005distribution,dalton2013exogenic}. This jovian plasma contains a high abundance of sulphur ions originating from Io's volcanos, which combines with the \ce{H2O} present on Europa's surface to produce the detected \ce{H2SO4.nH2O} \cite{carlson1999sulfuric}.

In addition, a number of hydrated salts have been proposed to explain the remainder of the non-ice contamination on Europa's surface, with this non-ice material likely consisting of a combination of hydrated sulphuric acid and salts \cite{carlson2009europa}. \citet{mccord1998salts}, \citet{mccord1999hydrated} and \citet{mccord2002brines} suggested that some mixture of magnesium and sodium sulphates and potentially sodium carbonates could explain the salts present in the NIMS spectra of Europa's trailing hemisphere. Salts were found to be concentrated in young lineae and chaos terrain, while the mixture of different salts appeared constant across the observed area, suggesting these salts may originate in the sub-surface ocean. Potential resurfacing mechanisms include cryovolcanic plumes, effusive flow from cracks in the crust and lag deposits from tectonic heating \cite{carlson2009europa}.

More recent advances in telescope optics have enabled studies using ground-based observatories to map compositional contrasts on Europa using at higher spectral resolution than Galileo/NIMS ($R \sim 60$). \citet{brown2013salts} used Keck/OSIRIS (\SIrange{1.4}{1.8}{\micro\m} and \SIrange{1.956}{2.381}{\micro\m}, $R\sim2000$, $\sim$\SI{100}{km/px}) to identify a small \SI{2.07}{\micro m} absorption feature on Europa's trailing hemisphere caused by magnesium sulphate salts. The apparent correlation of this feature with exogenic radiation products suggested that magnesium sulphate is itself a radiation product, rather than a completely endogenic constituent of Europa's sub-surface ocean. Observations with VLT/SINFONI (\SIrange{1.452}{2.447}{\micro\m}, $R=1500$, $\sim$\SI{70}{km/px}) \cite{ligier2016vlt} however suggested that magnesium chlorinated salts provide better fits to the overall spectrum than sulphate salts. These salts were found to be correlated with geological units, suggesting they have an endogenous origin from the sub-surface ocean. \citet{trumbo2019sodium} found that the \SI{450}{nm} sodium chloride absorption is strongly correlated with chaos terrain on Europa's leading hemisphere, again suggesting an endogenous origin with chlorinated salts in the sub-surface ocean.

The Juno mission has enabled spectroscopic observations with the JIRAM spectrometer (\SIrange{2}{5}{\micro\m}) from within the Jupiter system. Studies into \SI{>2}{\micro\m} water ice absorption bands suggest a mix of amorphous and crystalline ice with grain sizes ranging from tens to hundreds of microns \cite{filacchione2019serendipitous, mishra2021bayesian} and the temperature dependent reflectance peak around \SI{3.6}{\micro\m} suggests a maximum ice temperature of \SI{132}{K} \cite{filacchione2019serendipitous}.

None of the previous ground-based studies and very few of the NIMS observations cover the J band (\SI{<1.4}{\micro\m}) spectral range that is covered by SPHERE. This wavelength range covers spectral features such as water ice absorptions around \SI{1.1}{\micro\m} and \SI{1.3}{\micro\m} and various hydrated salt absorption bands around \SI{1.2}{\micro\m} (see \autoref{sec:spectral-library}). For ground-based observations, there is also a trade-off between spectral and spatial resolution, with the previous studies using instruments which offer higher spectral resolution at the expense of spatial resolution. SPHERE ($R \sim 30$, $\sim$\SI{25}{km/px}), on the other hand, offers a much higher spatial resolution, complementing these previous studies.

The icy Galilean moons, (Europa, Ganymede and Callisto), are due to be studied in the coming decade by ESA's \textit{Jupiter Icy Moons Explorer} (JUICE) and NASA's \textit{Europa Clipper}. JUICE will carry out flybys of Callisto and Europa, and will then orbit Ganymede, providing detailed global mapping data of the entire moon \cite{grasset2013jupiter}. Europa Clipper will also carry out flybys of Europa, Ganymede and Callisto, and will have a specific focus on Europa, performing 45 flybys of the moon to produce effectively global data coverage. The modelling of SPHERE and NIMS observations in this paper is a precursor to the analysis that will be possible with the MISE and MAJIS spectrometers on Europa Clipper and JUICE respectively.

In this paper, we will discuss our reduction and analysis of near-infrared reflected sunlight observations of Europa using the SPHERE instrument on the ground-based Very Large Telescope, VLT/SPHERE \cite{beuzit2019sphere}. These observations will be used to analyse the composition of the Europa's surface, in relation to the physical and chemical processes shaping these worlds. We will demonstrate that ground-based observations are capable of reproducing and extending those from visiting spacecraft, thereby providing key support for future missions.

\section{Observations and data reduction}
\label{sec:reduction}

\subsection{VLT/SPHERE}
\begin{figure*}
	\centering
	\includegraphics[width=\linewidth]{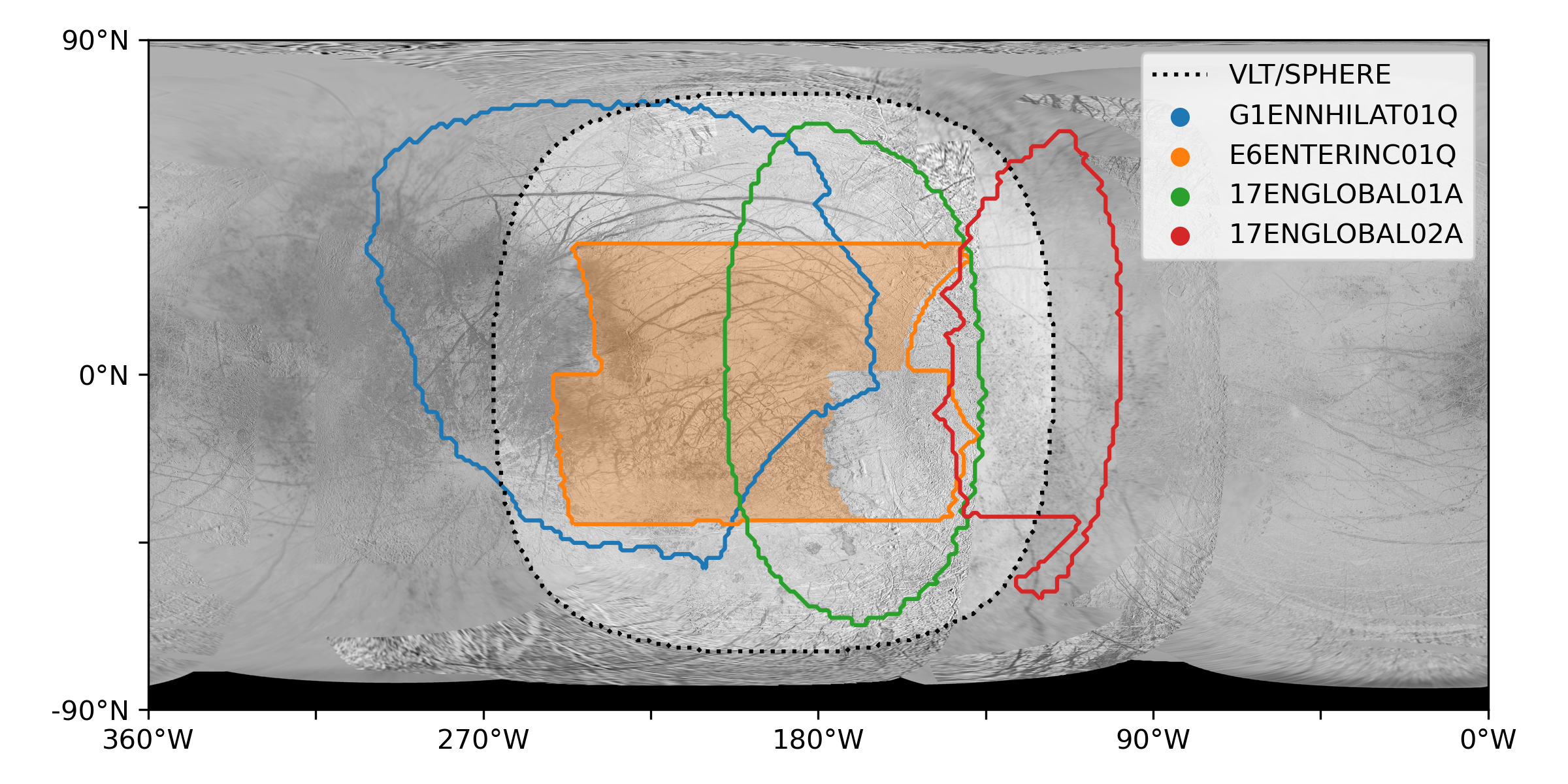}
	\caption{Map of spatial coverage of the datasets used in this study. The black dotted line shows the useful spatial coverage of the VLT/SPHERE observation. The coloured outlines show the total area observed in each NIMS dataset (see \autoref{tab:nims-observation-log}). The shaded orange region shows the area of the `E6' observation where NIMS has full spectral coverage of the SPHERE wavelength range, which is used for full spectral modelling. Regions with extreme emission and incidence angles ($>\ang{75}$) have large residual photometric errors, so are not used in this study or shown on this map. The background visible light reference image of Europa is from \citet{becker2013europa}. \ang{0}W is the sub-jovian longitude and \ang{180}W is the anti-jovian longitude.
		\label{fig:nims-selected}}
\end{figure*}
SPHERE (Spectro-Polarimetric High-contrast Exoplanet REsearch) \cite{beuzit2019sphere} is an instrument on the 8-m Very Large Telescope (VLT) UT3, located at the European Southern Observatory's Paranal Observatory in Chile. It was designed primarily as an exoplanet imager, using adaptive optics to provide stable high-spatial-resolution observations, enabling direct imaging of exoplanets orbiting their host stars. This high spatial resolution can also be applied to other classes of objects, such as the solar system targets like Europa.

Observations of Europa were taken during SPHERE science verification in December 2014, as summarised in \autoref{tab:observation-log} and shown in \autoref{fig:nims-selected}. The observation block consists of multiple Europa observations, and then a single calibration star observation, taken immediately after the Europa observations. This ensured that the atmospheric conditions for the science and calibration observations were as similar as possible. We used SPHERE in \verb|IRDIFS_EXT| mode, allowing simultaneous imaging with the Integral Field Spectrograph (IFS) and Infrared Differential Imaging Spectrometer (IRDIS) sub-systems of the SPHERE instrument. This allows simultaneous spectroscopic observations covering \SIrange{0.95}{1.65}{\micro\m} (J and H bands) and dual band imaging at \SI{2.1}{\micro\m} and \SI{2.251}{\micro\m} (K band).

\begin{deluxetable*}{ccccc}
	\tablecaption{VLT/SPHERE observation log for Europa and calibration star observations. \label{tab:observation-log}}
	\tablehead{\colhead{Time (UTC)} & \colhead{Target} & \colhead{Sub-observer point} & \colhead{Sub-solar point} & \colhead{Phase angle}}
	\startdata
	2014-12-09T07:51:21  & Europa & \coords{192.1}{0.2}& \coords{201.8}{0.6}& \ang{9.7160}\\
	2014-12-09T07:52:54  & Europa & \coords{192.2}{0.2}& \coords{202.0}{0.6}& \ang{9.7159}\\
	2014-12-09T07:54:27 & Europa & \coords{192.3}{0.2}& \coords{202.0}{0.6}& \ang{9.7158}\\
	2014-12-09T07:56:00 & Europa & \coords{192.4}{0.2}& \coords{202.2}{0.6}& \ang{9.7157}\\
	2014-12-09T07:57:34 & Europa & \coords{192.6}{0.2}& \coords{202.3}{0.6}& \ang{9.7156}\\
	2014-12-09T08:09:07 & BD+17 2101 & \multicolumn{3}{c}{(calibration star)}\\
	\enddata
\end{deluxetable*}

The Integral Field Spectrograph (IFS) is a relatively new tool in astronomy, enabling both spatial and spectral information to be recorded simultaneously on the same detector. The IFS on SPHERE \cite{claudi2008sphere, mesa2015performance} produces image cubes with 38 wavelength channels from \SIrange{0.95}{1.65}{\micro\m}, giving a low spectral resolution of $R = \lambda/\Delta\lambda \sim 30$. It has a high spatial resolution, with a pixel size of \SI{7.46}{mas/px}, corresponding to \SI{\sim25}{km/px} at Jupiter. Accounting for diffraction, this allows features \SI{\sim150}{km} across to be resolved. The magnitude of the diffraction is calculated by measuring the full width at half maximum of the point spread of the calibration star, observed through VLT/SPHERE (\SI{\approx6}{px} at \SI{0.95}{\micro\m}).

The IFS operates by using a lenslet array that focusses the image into a series of \textit{spaxels} (spatial pixels), spots in the focal plane, each of which is from a different area of the observed field. These spaxels are used as the input into a dispersive spectrograph that transforms each spaxel into a full spectrum for its respective part of the image. These multiple spectra (one from each spaxel) are imaged by a single detector, recording both the spatial and spectral information from the observation. In the reduction process, each spectrum is fitted and transformed into the spectral dimension of the output image cube so that the final image cube has one slice for each measured wavelength \cite{claudi2008sphere}.

IRDIS \cite{dohlen2008infra, vigan2010photometric} produces simultaneous imaging through two filters, producing two images on separate parts of the same detector. Different filter combinations are available so that contrast can maximised depending on the spectral features of individual targets. Our data uses the \verb|DB_K12| filter set that have transmissions centred at \SI{2.1}{\micro\m} for the K1 filter and \SI{2.251}{\micro\m} for the K2 filter, with filter widths of \SI{0.051}{\micro\m} and \SI{0.055}{\micro\m} respectively, meaning IRDIS is sensitive to water ice absorption at \SI{2}{\micro\m}.

\subsection{Data reduction}
The data reduction routine transforms the raw observations into cleaned and calibrated data sets that are ready for scientific use. The reductions account for detector properties, atmospheric conditions and observation conditions to ensure observations are calibrated and comparable.

The raw IRDIS data consist of FITS image files containing two images side-by-side on the same detector, one from each filter. The IRDIS reduction process is written in Python and is as follows:

\begin{enumerate}
	\item Perform instrument calibration by subtracting dark frame from observations and dividing by flat frame.
	\item Split the image into two images, one for each filter.
	\item Clean the images by `despiking' to remove extreme bright or dark pixels.
	\item Sum images from each observation's detector integrations into a single image for each filter in each observation.
\end{enumerate}

After reduction, the IRDIS images can be corrected and mapped in a similar way to the IFS data (below).

The raw IFS data consists of FITS files containing individual spectra for each spaxel, arranged across the detector. Our reduction process consists of a series of steps to ultimately transform these individual spectra into a mapped spectral cube for the observed target. The reduction routine is written in Python, and makes use of EsoRex, the ESO Recipe Execution Tool that provides standard reduction pipelines for ESO instruments. An overview of our IFS data reduction routine is:

\begin{enumerate}
	\item Clean raw images by replacing known bad pixels with interpolated values.
	\item Run EsoRex reduction to identify spectral locations on detector, perform instrument calibrations (using detector flat and dark frames) and generate image cubes. This produces an image cube for each individual detector integration in each individual observation.
	\item Sum image cubes from each observation's detector integration into a single cube for each observation. This improves signal-to-noise ratio and improves processing efficiency, as the number of files to process is decreased significantly.
	\item Clean the image cubes by `despiking' (replacing extreme bright or dark pixels) and `destriping' to remove a prominent banding pattern (see \autoref{sec:destriping}).
	\item Calibrate the image cubes to remove telluric contamination and calculate reflectance spectra, using the calibration star observation (see \autoref{sec:spectral-calibration}).
	\item Map the observations by performing a photometric correction and transforming the images to an equirectangular map projection (see \autoref{sec:mapping}).
	\item Combine the maps from individual observations into a single mapped cube for each target. This gives a complete spectrum for each observed point on the map.
\end{enumerate}

\subsubsection{Image cleaning}
\label{sec:destriping}
\begin{figure*}
	\centering
	\includegraphics[width=0.75\linewidth]{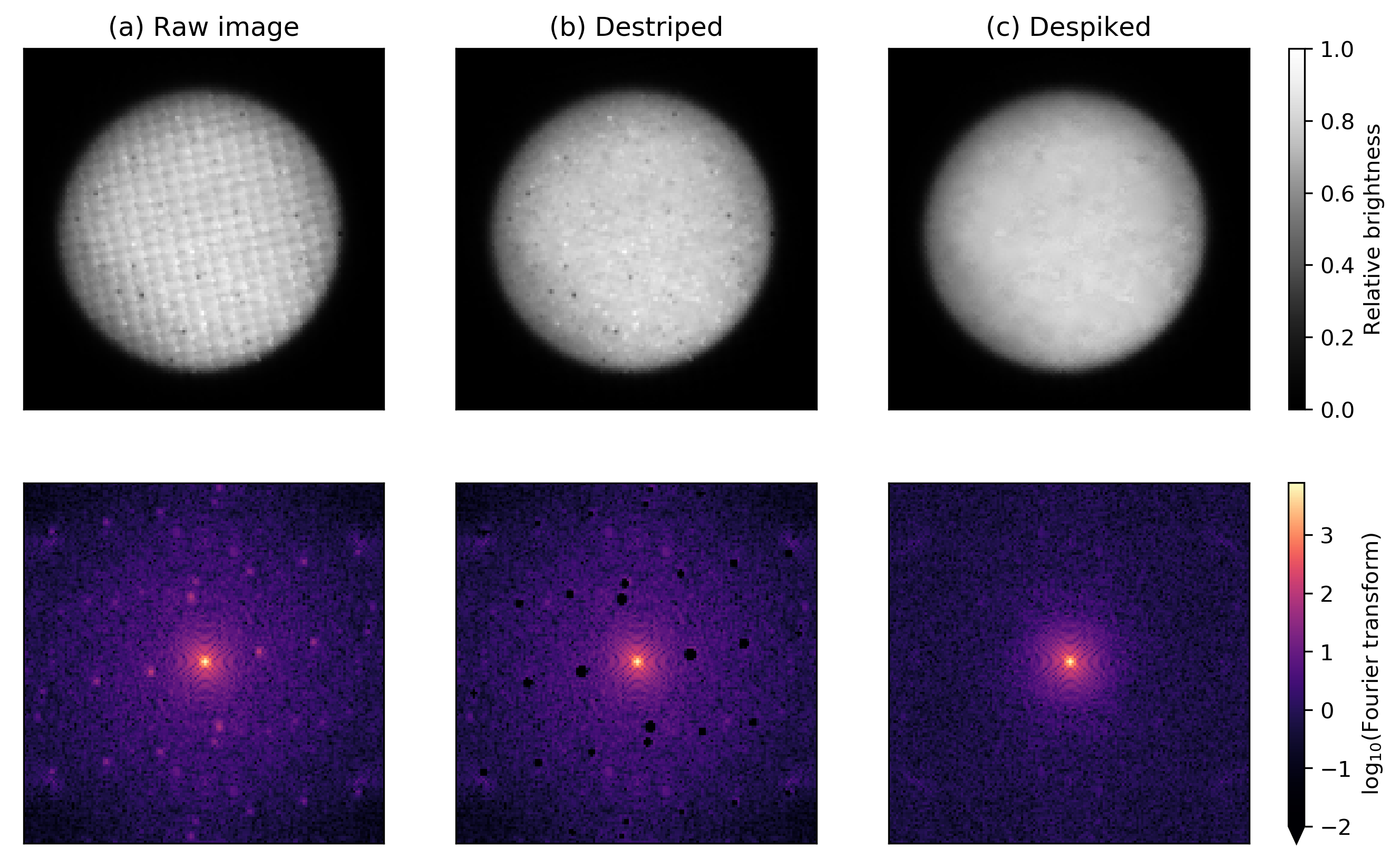}
	\caption{Image cleaning process for IFS data. The top images show observations of Europa at \SI{1.37}{\micro\m}, the wavelength where the striping pattern is the worst. The bottom images show the observations' corresponding Fourier Transform magnitudes on a logarithmic scale. The frequencies that cause the striping pattern in the raw image (the bright dots away from the centre of the Fourier transform) are set to zero to remove the artificial pattern. The image is then despiked to remove any extreme bright or dark pixels, which are replaced by interpolated values. \label{fig:ifs-cleaning}}
\end{figure*}

A strong regular banding pattern (see \autoref{fig:ifs-cleaning}a) remains after the EsoRex reduction for extended sources such as Europa, so we developed a method using a Fourier transform filter to identify and remove this artificial pattern. This pattern is likely to be produced from cross-talk between different lenslets not being fully accounted for in the EsoRex step in the reduction pipeline. The pattern is regular and fixed in position for different wavelengths and observations, so it can be systematically removed using a Fourier transform filter to remove the specific frequencies that generate the pattern. The strength of the pattern is wavelength dependent, and is strongest at \SI{\sim1.35}{\micro\m}.

\begin{figure}
	\centering
	\includegraphics{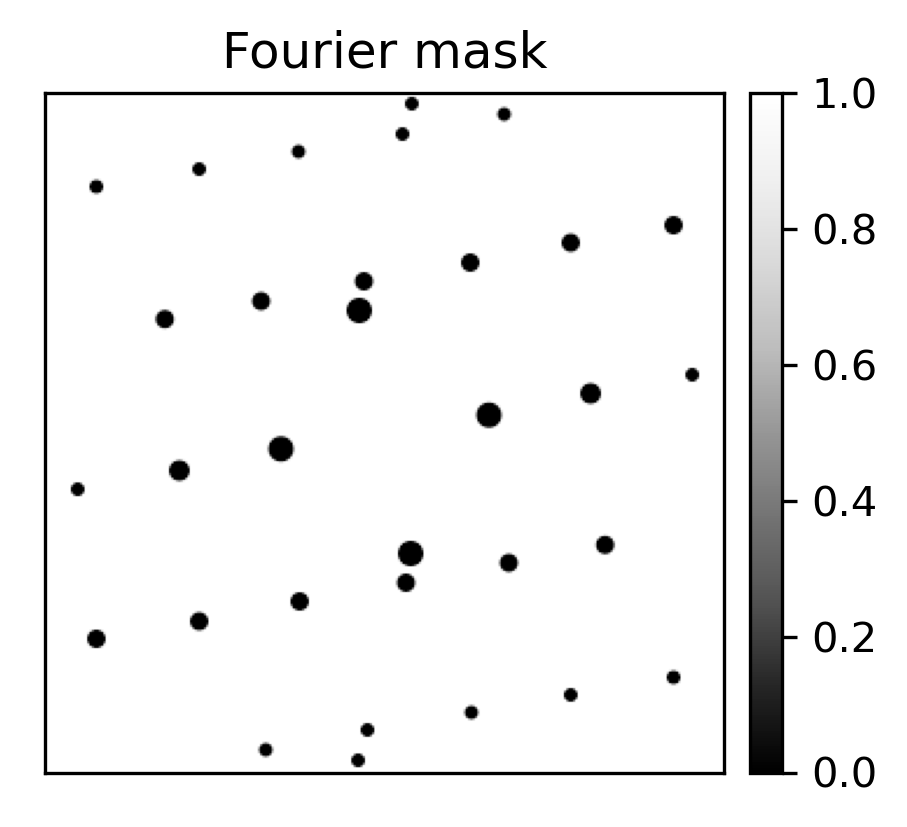}
	\caption{Mask used to remove unwanted frequencies from Fourier transform of IFS data. The mask is multiplied with the Fourier transform (see \autoref{eq:destriping}), so white areas (=1) are kept and black areas (=0) are removed. \label{fig:destripe-mask}}
\end{figure}

To remove this pattern, its specific frequencies were identified by analysing the Fourier transform of IFS images. These frequencies are removed from the image by converting the image to Fourier space, setting the frequencies to zero, then converting back to image space using

\begin{equation}
	\text{Destriped} = \mathcal{F}^{-1}\left\{\text{Mask} \times \mathcal{F}\left\{\text{Image}\right\}\right\}
	\label{eq:destriping}
\end{equation}

where $\mathcal{F}$ is the Fourier transform (image domain to frequency domain) and $\mathcal{F}^{-1}$ is the inverse Fourier transform (frequency domain to image domain). The mask used to remove the unwanted frequencies is shown in \autoref{fig:destripe-mask}.

The destriping routine removes the pattern for wavelengths where it occurs, and has negligible effect for wavelengths where the pattern does not occur. Therefore, it can be safely applied to every wavelength of an image cube to avoid introducing any wavelength-dependent systematic errors.

Images are `despiked' by calculating the mean and variance of all pixels within a 2~px radius of a given pixel. If the pixel value is greater than two standard deviations from the mean of the surrounding pixels, it is assumed to be a `bad' pixel, and its value is replaced with the mean value. This removes extreme spurious values whilst preventing loss of actual data, as diffraction means that any variation in the data would be seen across multiple pixels. The despiking routine is run repeatedly until there are no bad pixels to replace. It is based on the \verb|sigma_filter| routine in the IDL Astronomy User's Library \cite{idlastro}.

See \autoref{fig:ifs-cleaning} for a summary of this image cleaning process.

\subsubsection{Spectral calibration}
\label{sec:spectral-calibration}
The spectral calibration calculates the reflectance spectrum of the observed moon from the measured flux. The calibration star observation is used to radiometrically calibrate the spectrum, removing any telluric contamination in the process. This process, based on the telluric correction step described in \textcite{ligier2016vlt}, uses the known spectrum of the calibration star and the known solar spectrum to calculate the reflectance

\begin{equation}
	R(\lambda) = \frac{F_\text{moon}(\lambda) - B(\lambda)}{F_\text{star}(\lambda)}
	\times \frac{T_\text{star}}{T_\text{moon}}
	\times \frac{1}{\Omega \times 10^{\frac{m_\text{star}(\lambda) - m_\text{sun}(\lambda)}{2.5}}}
	\label{eq:spectral-calibration}
\end{equation}

where $R(\lambda)$ is the reflectance spectrum of Europa. $F_\text{moon}$ and $F_\text{star}$ are the measured fluxes from Europa and the calibration star respectively, $B(\lambda)$ is the background flux, $T$ is the integration time of an observation, $\Omega$ is the solid angle subtended by each detector pixel, $m_\text{star}(\lambda)$ is the apparent magnitude of the calibration star at Earth \cite{cutri2003vizier} and $m_\text{sun}(\lambda)$ is the apparent magnitude of the Sun at Europa \cite{blanton2007k}.

The stellar and background fluxes are calculated by summing the flux in two circular apertures, both centred on the calibration star that have radii calculated from the size of the Airy disc. The star aperture is chosen so that the first airy diffraction maximum is contained within the aperture while still minimising the background flux within the aperture. The background flux is summed from a \SI{10}{px} wide annular aperture around the same calibration star which reduces the effect on the calibration of any slight variation in sensitivity across the detector. The average spectral slope of the SPHERE data is finally adjusted to be equal to the spectral slope from the NIMS data to remove any slight variations caused by differing spectral slopes from the calibration star spectrum and the solar spectrum. This removes the need to allow the continuum spectral slope to vary as part of the spectral fitting process as in \citet{ligier2016vlt}.

\subsubsection{Mapping}
\label{sec:mapping}
In order to map the observations, the exact location, size and orientation of the moon's disc in the observed image must be known. The size of the moon (i.e. its pixel radius) and its orientation (i.e. the angle between its north pole and the image vertical) can be calculated from ephemeris data and the known plate scale and orientation of the detector.

The pointing information for the telescope (RA and Dec) are not accurate enough to use for determining the location of the moon's disc in the image for mapping purposes, as accurate mapping and photometric correction requires the disc location to be known to $\lesssim\SI{1}{px}$ for the best results. Therefore, the disc location is calculated from the image itself. This is done by applying a threshold to the image (such that all dark pixels are 0 and bright pixels are 1), and then calculating the centroid of this image (its centre of brightness). The threshold step ensures that any variations in brightness across the moon do not significantly bias the results, as all pixels within the disc of the moon will have the same value. This method identifies the centre of the disc within a few pixels.

The calculated disc location is then corrected to reduce errors caused by the moon's phase. This is done by creating a photometric model of the target and calculating its apparent position through the same routine. The difference between the actual centre of the photometric model and its calculated centre is then used to correct the calculated centre of the observation. This step ensures the dark limb of a target observed at non-zero phase angles does not affect the calculation of the its centre. For our Europa data, observed at a phase angle of \ang{\sim10}, this step corrects the disc location by \SI{\sim2}{px}.

Once the position of the disc is known, the latitude, longitude, incidence angle and emission angle are calculated for each pixel using ephemeris data for the target and the rotation information for the telescope. The incidence and emission angles are used to photometrically correct the image (\autoref{sec:photometric-correction}). Finally, the image is transformed to an equirectangular map projection using the calculated coordinates for each pixel. Each wavelength of each individual observation is mapped separately to ensure there were no errors introduced by Europa's rotation between observations or any wavelength dependent drifts.

\subsubsection{Photometric correction}
\label{sec:photometric-correction}
\begin{figure*}
	\centering
	\includegraphics[width=0.75\linewidth]{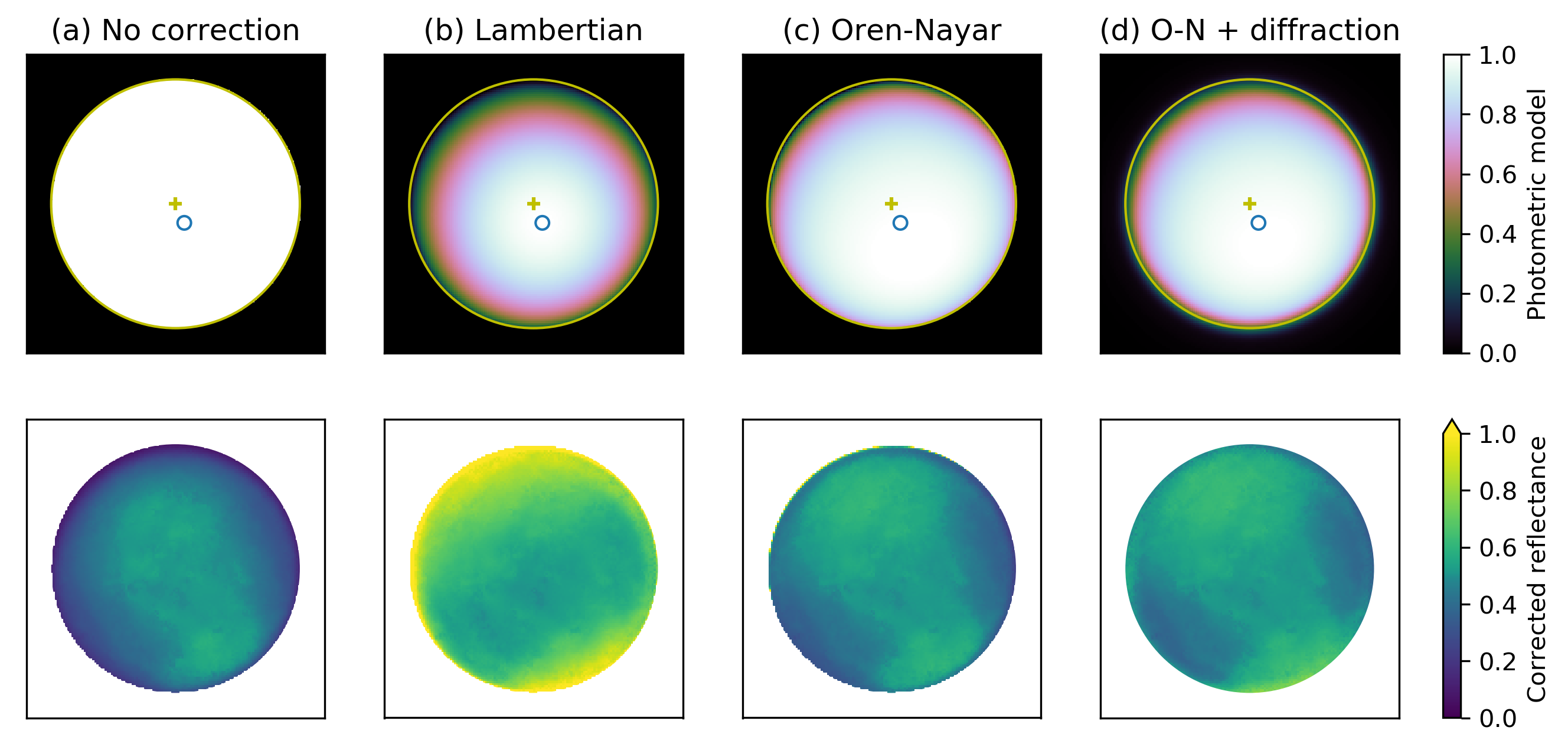}
	\caption{Summary of different photometric models (top) and images of Europa at \SI{1.65}{\micro\m} with the photometric correction applied (bottom). The sub-solar point is marked by the blue circle, the sub-observer point by the yellow cross and the outline of the modelled disc is given by the large yellow circle. This Europa observation has a solar phase angle of \ang{9.7}, and the final photometric model (d), combines the Oren-Nayar model (with surface roughness $\sigma = \ang{33}$) and a diffraction model (with $r_{\text{airy}} = \SI{10}{px}$).}
	\label{fig:photometric-correction}
\end{figure*}
\begin{figure}
	\centering
	\includegraphics[width=\linewidth]{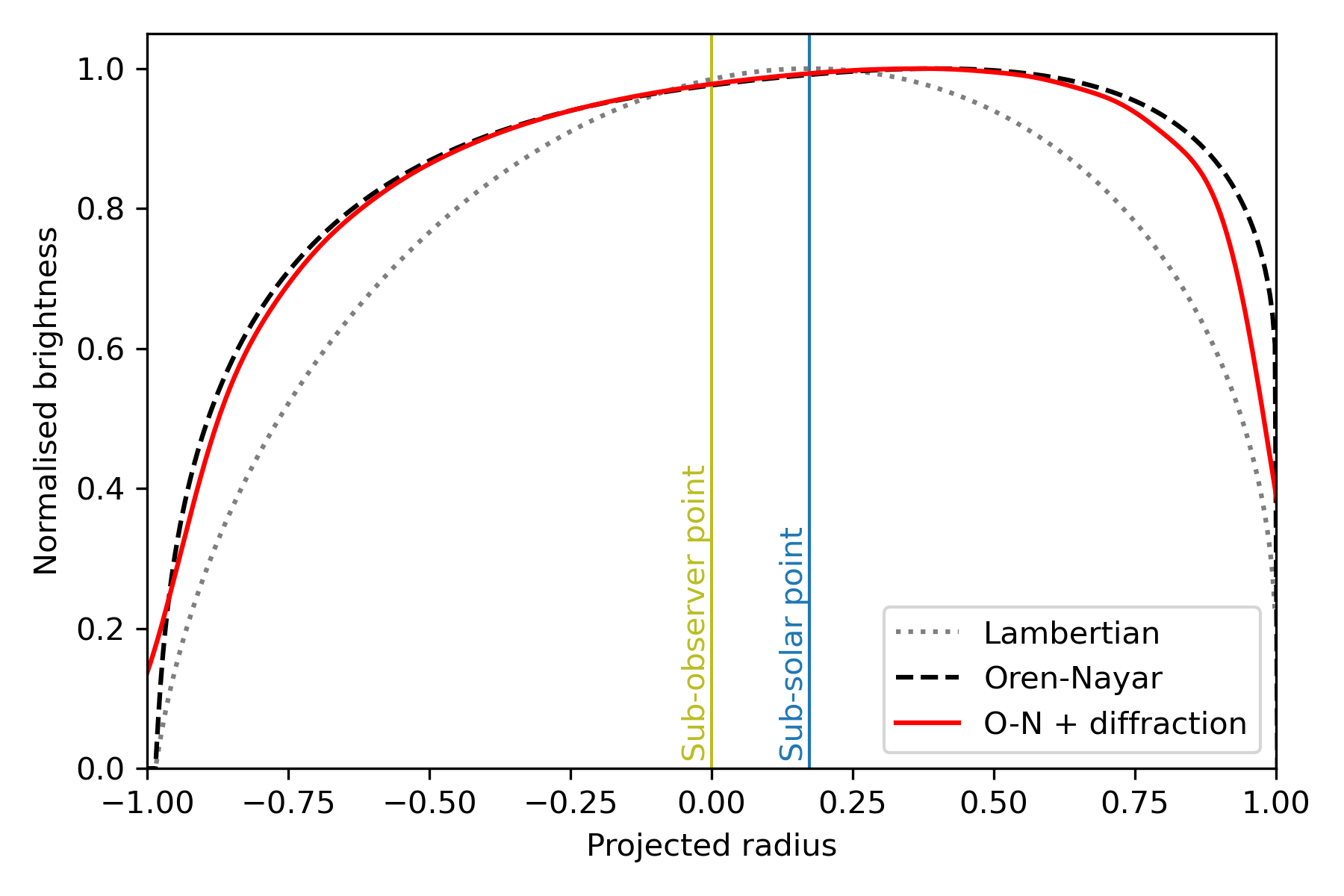}
	\caption{Brightness curves for different photometric models of Europa with a solar phase angle of \ang{9.7}. The final photometric model (`O-N + diffraction') has surface roughness $\sigma = \ang{33}$ and $r_{\text{airy}} = \SI{10}{px}$.
		\label{fig:photometric-curves}}
\end{figure}

A sphere of uniform albedo illuminated by a point-like source (i.e. the sun) appears to have varying brightness, with brightness dropping off towards the edge of the disc (see \autoref{fig:photometric-correction}a). The photometric effect must be accounted for to correct the reflectance for areas of an observed disc away from the centre. The correction takes the form

\begin{equation}
	\text{Corrected image} = \frac{\text{Image}}{\text{Photometric model}}
\end{equation}

where the photometric model gives the ratio of the brightness of the disc to the brightness of a surface viewed and illuminated at a normal incidence angle. Therefore, the values of the corrected image give the reflectance each location would have if it was both the sub-solar point and the sub-observer point (i.e. where the incidence and emission angles are zero).

The simplest photometric correction assumes a Lambertian surface - a surface that is a perfect diffuse scatterer and therefore appears equally bright from all directions \cite{hapke1993book, lambert1760photometria}. Therefore, the brightness is simply proportional to the projected area of the surface to the illumination source, $\cos(i)$, where $i$ is the angle of the source from the surface normal. The Lambertian correction improves on the basic images, but accumulates significant errors towards the edge of the disc, overcompensating for the limb darkening and producing unphysical reflectance values ($>1$) (see \autoref{fig:photometric-correction}b).

Therefore, we used the Oren-Nayar reflectance model \cite{oren1994generalization}, which modifies on the Lambertian model by accounting for surface roughness, a significant improvement on the implicit assumption of a perfectly smooth surface in the Lambertian model. The Oren-Nayar model assumes the surface is composed of a series of V-shaped cavities, with facets that are themselves Lambertian scatterers. The cavities are modelled to have a Gaussian distribution of slopes, with a mean slope of $\mu=0$ and a standard deviation of $\sigma$ that parametrises the surface roughness. The Oren-Nayar model reduces to standard Lambertian scattering for $\sigma = 0$.

For a surface roughness $\sigma$ and albedo $\rho$, the modelled Oren-Nayar surface brightness is

\begin{align}
	f_r = {} & \cos(i)\left(L_1 + L_2\right)                                                                                                   \\
	L_1 = {} & \frac{\rho}{\pi}(C_1 + \cos(\phi)\tan(\beta)C_2 \nonumber                                                                       \\
	         & {} +(1 - |\cos(\phi)|)\tan\left(\frac{\alpha+\beta}{2}\right)C_3)                                                               \\
	L_2 = {} & 0.17\frac{\rho^2}{\pi}\left(\frac{\sigma^2}{\sigma^2 + 0.13}\right)\left(1 - \cos(\phi)\left(\frac{2\beta}{\pi}\right)^2\right)
\end{align}

where $(i, \phi_i)$ and $(e, \phi_e)$ are the polar and azimuth angles of the incident and emitted rays respectively. $L_1$ is the contribution from single scattering and $L_2$ is the contribution from multiple scattering with

\begin{align}
	\phi   & = \phi_e - \phi_i  \\
	\alpha & = \text{Max}(e, i) \\
	\beta  & = \text{Min}(e, i)
\end{align}

The single scattering coefficients $C_n$ are dependent on the surface roughness $\sigma$ and the viewing geometry,

\begin{align}
	C_1 & = 1 - 0.5\frac{\sigma^2}{\sigma^2 + 0.33}                                                                                             \\
	C_2 & =
	\begin{cases}
		0.45\frac{\sigma^2}{\sigma^2 + 0.09}\sin(\alpha)                                                  & \text{if}\ \cos(\phi) \geq 0 \\
		0.45\frac{\sigma^2}{\sigma^2 + 0.09}\left(\sin(\alpha) - \left(\frac{2\beta}{\pi}\right)^3\right) & \text{otherwise}
	\end{cases} \\
	C_3 & = 0.125\left(\frac{\sigma^2}{\sigma^2 + 0.09}\right)\left(\frac{4\alpha\beta}{\pi^2}\right)^2
\end{align}

Incorporating roughness increases the modelled brightness at the edge of the disc when compared to the Lambertian model, as shown in \autoref{fig:photometric-curves}. The surface roughness parameter $\sigma$ was selected by photometrically correcting the observation with a range of $\sigma$ values and visually comparing the corrected brightness at the limb and towards centre of Europa's disk. Large values of $\sigma$ produce undercorrected observations where the limb is too dark (e.g. \autoref{fig:photometric-correction}a) and small values of $\sigma$ produce overcorrected observations where the limb is too bright (e.g. \autoref{fig:photometric-correction}b). $\sigma = \ang{33}$ (\autoref{fig:photometric-correction}c) was found to produce the minimum brightness variation so was selected as our surface roughness parameter. Some small residual photometric errors are still present at the very edge of the disk with $\sigma = \ang{33}$, so we conservatively constrain the region of useful data to $e<\ang{75}$.

The only other free parameter required by the model is the surface's albedo. Variation in the albedo has a very negligible effect on the modelled photometry, so precise fitting of this value was not required and a value of $\rho=0.5$ was used.

To account for the diffraction of the telescope optics, the photometric model is convolved with an airy disc model before being used to correct the image. The radius of the airy disc is proportional to the wavelength of the specific image being corrected, with the radius set to $r_{\text{airy}} = \SI{10}{px}$ for \SI{1.65}{\micro\m}, as measured from observations of the calibration star (see \autoref{fig:photometric-correction}d for final photometric model and corrected observation).

This photometric correction allows accurate correction of the images to emission angles $e>\ang{70}$, corresponding to $\sim90\%$ of the observed disc. This is higher than previous studies using Lambertian models that typically extend to \ang{50} - \ang{60} \cite{brown2013salts, grundy2007new, ligier2016vlt}. Improvements beyond \ang{\sim75} are unlikely to provide much more useful data as the extreme viewing angles at the edge of the disc produce very low spatial resolutions once these areas are mapped. Therefore, we conservatively use $e=\ang{75}$ as the limit of the reliable mapped region of the data.

\subsection{Galileo/NIMS}
The Galileo orbiter was launched in 1989, and orbited Jupiter from 1995 to 2003. Galileo's orbit included a series of flybys of Europa, enabling detailed imaging and mapping of Europa's surface.

The Near-Infrared Mapping Spectrometer (NIMS) \cite{carlson1992near}, instrument on Galileo performed spectroscopy from \SIrange{0.7}{5.2}{\micro\m} with a spectral resolution of $R = \lambda/\Delta\lambda \sim 60$. During flybys, a series of scans by the NIMS instrument would map the infrared spectra of Europa's surface beneath the spacecraft. The spatial resolution of the mapping was dependent on the distance between Galileo and Europa, with spatial resolutions \SI{<1}{km/px} for the closest flybys. However, there are also large regions of Europa's surface with no NIMS data or only very low spatial resolution data(\SI{>500}{km/px}).

The NIMS instrument used different detectors to cover different wavelength ranges, one of which (detector 3, covering \SIrange{0.99}{1.26}{\micro\m}) failed early in the Galileo mission, during Galileo's 4th orbit of Jupiter, so there are no Europa observations with full spectral coverage from 1997 onwards. Therefore, there are large regions of Europa with no observations or only very low spatial resolution observations at these wavelengths. The SPHERE IFS wavelength range includes this spectral range so SPHERE can fill this missing gap in the near-infrared spectra of the Galilean moons.

\begin{deluxetable}{lcc}
	\tablecaption{Galileo/NIMS datasets used in this study. \label{tab:nims-observation-log}}
	\tablehead{\colhead{Name} & \colhead{Date} & \colhead{Spatial Resolution}}
	\startdata
	G1ENNHILAT01Q & 1996-06-28 & \SI{\sim 80}{km/px} \\
	E6ENTERINC01Q & 1997-02-20 & \SI{\sim 50}{km/px} \\
	17ENGLOBAL01A & 1998-09-25 & \SI{\sim 50}{km/px} \\
	17ENGLOBAL02A & 1998-09-26 & \SI{\sim 50}{km/px} \\
	\enddata
\end{deluxetable}

The NIMS datasets given in \autoref{tab:nims-observation-log} were used in this study. These datasets were taken at a low solar phase angle, and collectively cover Europa's anti-jovian hemisphere, allowing direct comparison to our SPHERE observations of the same region. The datasets 17ENGLOBAL01A and 17ENGLOBAL02A were taken after the failure of NIMS detector 3, so lack the \SIrange{0.99}{1.26}{\micro\m} spectral range.

All NIMS observations of Europa were downloaded from the NASA Planetary Data System, and reduced using a pipeline similar to that for our ground based observations. The I/F NIMS data cubes were processed using ISIS3 \cite{isis} to calculate the viewing angles (solar incidence angle, emission angle and phase angle) and coordinates (latitude and longitude) for each observed pixel in the NIMS images. These angles and coordinates were used to map and photometrically correct the datasets using the same mapping and photometric correction code as the SPHERE dataset.

For our analysis, we selected NIMS observations taken at a low phase angle with a relatively high spatial resolution ($\lesssim$\SI{100}{km/px}) and similar spatial coverage to our SPHERE observation (spatial resolution $\sim$\SI{150}{km}), allowing direct comparison between the two datasets.

\section{Water ice absorption bands}
\label{sec:discussion}
\begin{figure*}
	\centering
	\includegraphics[width=\linewidth]{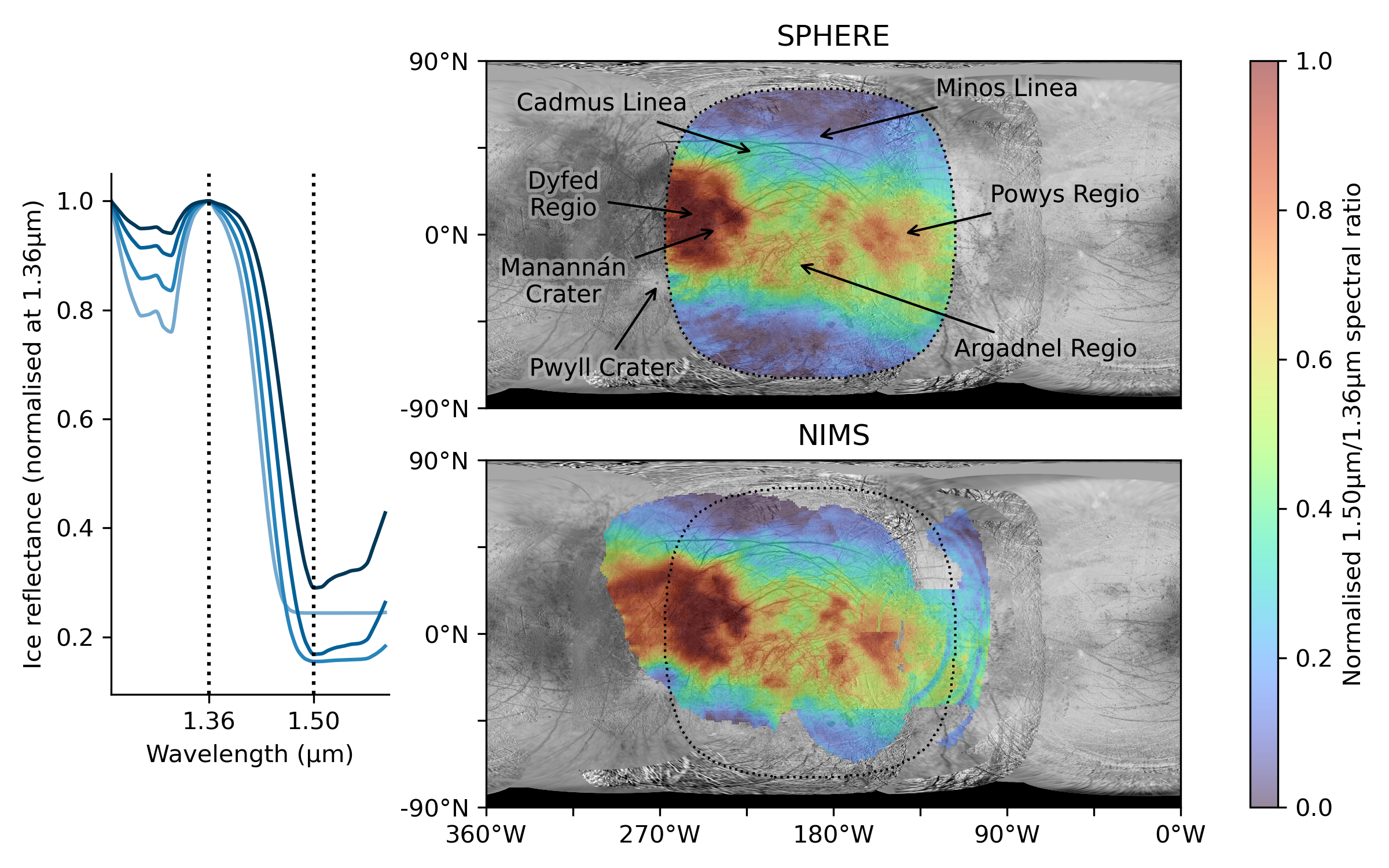}
	\caption{Normalised $1.50/\SI{1.36}{\micro\m}$ spectral ratio for SPHERE and NIMS observations. The NIMS data are a combination of the datasets in \autoref{tab:nims-observation-log}; in regions where the observations overlap, the median reflectance value is used. Blue areas in the map indicate stronger absorption at \SI{1.50}{\micro m}, generally implying a higher water ice abundance and red areas indicate reduced water ice abundance. The graph on the left shows reflectance spectra for a range of water ice grain sizes (darker lines indicate smaller ice grains, see \autoref{fig:example-endmembers}) normalised to unity at \SI{1.36}{\micro\m}. The choice of ratio wavelengths measures the full relative depth of the \SI{1.5}{\micro\m} absorption and produces a map ratio which is mainly affected by ice abundance (rather than grain size). \label{fig:1.5-band-map}}
\end{figure*}
\begin{figure*}
	\centering
	\includegraphics[width=0.75\linewidth]{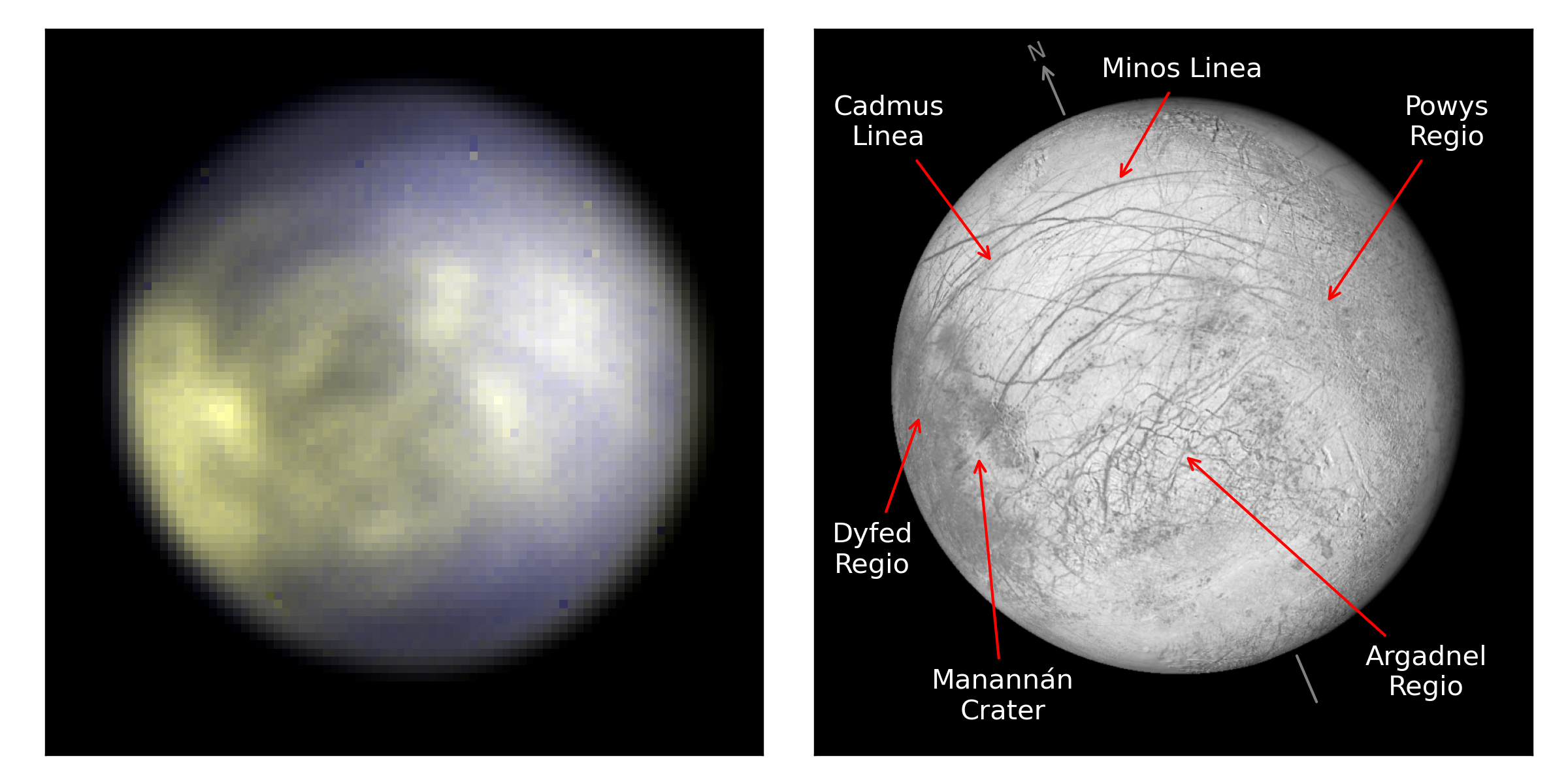}
	\caption{IRDIS two-colour observation of Europa (left), and simulated visible light reference image of Europa from \citet{becker2013europa}. The two-colour IRDIS observation is produced using the K1 filter at \SI{2.1}{\micro\m} (filter width \SI{0.051}{\micro\m}), shown in yellow, and the K2 filter at \SI{2.251}{\micro\m} (width \SI{0.055}{\micro\m}) in blue. Water ice has a broad absorption band around \SI{2}{\micro\m}, so the blue areas are icy and yellow areas dominated by non icy species.
		\label{fig:irdis-two_color}}
\end{figure*}

The most prominent features in Europa's near-IR spectra are a series of absorption bands caused by water ice \cite{greeley2004geology}. These bands can be used to provide a simple qualitative indication of the water ice distribution on the surface of icy bodies like Europa, with stronger water ice absorption corresponding to higher water ice abundance \cite{hapke1993book}. As Europa's crust is mainly composed of water ice \cite{greeley2004geology}, the inverse of the water ice distribution provides an indication of the spatial distribution of non-ice contaminants on Europa's surface.

\autoref{fig:1.5-band-map} shows the strength of the mapped \SI{1.5}{\micro\m} band in the SPHERE IFS and Galileo/NIMS datasets using the $1.50/\SI{1.36}{\micro\m}$ spectral ratio. The use of the ratio (rather than simply measuring the reflectance at \SI{1.50}{\micro\m}) means that variations in this relative band depth are mainly caused by variations in water ice abundance rather than varying grain size or ice crystallinity (see graph in \autoref{fig:1.5-band-map}). \autoref{fig:irdis-two_color} shows the strength of the \SI{2}{\micro\m} water ice band, as directly imaged by the two filters of the IRDIS instrument.

Both bands and instruments show consistent distributions, with the only major differences due to the higher spatial resolution of the NIMS dataset. Water ice appears more abundant at high latitudes and there is significant contamination of Europa's trailing hemisphere, particularly in Dyfed Regio. This contamination is centred on the trailing apex and is consistent with the bullseye distribution of exogenous plasma bombardment increasing the non-ice fraction of the trailing hemisphere at low latitudes \cite{carlson2005distribution}.

The shape of the contaminated regions follow the structures of geological units, such as the clear outlines of Dyfed Regio and Powys Regio. The effect of Manannán and Pwyll impact craters are visible due to the impacts exposing less contaminated ice leading to an increased water ice signature in and around the craters. Signatures of some of Europa's large lineae are also visible in both the SPHERE and NIMS data around Cadmus Linea, as an arc of lower ice content material from the north of Dyfed Regio. This is the first time that the lineae have been directly observed using a ground-based telescope, demonstrating the significant improvement in spatial resolution that SPHERE enables.

The Oren-Nayar photometric correction allows our mapping to extend to higher emission angles than typical for ground-based observations. Here, we assume data below an emission angle of $e=\ang{75}$ is useful, as above this angle the spatial resolution of the mapped data degrades significantly. This is an improvement on previous studies, which typically discard data above $e=\ang{50} \text{ to } \ang{60}$, allowing us to usefully observe and map higher latitudes with ground-based observations.

\section{Spectral modelling}
\label{sec:modelling}
\subsection{Linear unmixing}
The mapped spectral cubes are analysed by fitting to laboratory spectra from reference cryogenic libraries (described in \autoref{sec:spectral-library}). Our fitting routine uses linear spectral modelling to fit the observed spectra, where the modelled spectrum, $M$, is

\begin{equation}
	M(\lambda) = \sum_{i}w_i E_i(\lambda)\times S
	\label{eq:linear-spectral-model}
\end{equation}

where $w_i$ are the weights of the different endmembers $E_i$. The parameter $S$ accounts for any continuum spectral slope that may remain in the data. The weights $w_i$ give the fractional abundance of each endmember in the modelled spectrum, and are subject to the constraint

\begin{equation}
	\sum_{i} w_i = 1
	\label{eq:sum-to-one-constraint}
\end{equation}

which ensures the fractional abundance of the different endmembers sums to 100\% and

\begin{equation}
	0 \leq w_i \leq 1
\end{equation}

which ensures the individual abundances are all physically realistic values (0\% to 100\%).

The use of linear unmixing is generally valid for ground-based observations, as the relatively low spatial resolutions ($\sim 100$~km) mean that the observed spectrum for each pixel is naturally a linear combination of the spectra of different geological units within that pixel \cite{ligier2016vlt}. This method will not however account for any non-linear scattering that occurs, and may therefore be responsible for some small residual errors after fitting \cite{ligier2016vlt, shirley2016europa}.

\subsection{Markov Chain Monte Carlo modelling}
\label{sec:mcmc-fitting}
Markov Chain Monte Carlo (MCMC) simulates systems by sequentially randomly sampling a multidimensional parameter space. MCMC uses a series of `walkers' that sequentially model the system in a chain of different parameter values, ultimately building a posterior distribution of parameter values consistent with the observed data.

Bayes theorem \cite{bayes1763lii} states that the posterior probability of a set of parameter values $\vec{w}$ given an observation $\vec{o}$ can be calculated as

\begin{equation}
	P\left(\vec{w}|\vec{o}\right) \propto P\left(\vec{o}|\vec{w}\right) P\left(\vec{w}\right)
\end{equation}

where $P\left(\vec{o}|\vec{w}\right)$ gives the likelihood of measuring the observation $\vec{o}$ given the parameters $\vec{w}$. $P\left(\vec{w}\right)$ gives the prior probability of a set of parameter values and is used to include any information known before taking any observations (e.g. the constraint that abundances must sum to 100\%).

At a given iteration of the MCMC chain, a walker will have a specific set of free parameter values, which can be described as the vector $\vec{w}_n$. These parameter values can be related to the observed data using a cost function

\begin{equation}
	C(\vec{w}) \propto -\log\left(P\left(\vec{w}|\vec{o}\right)\right)
\end{equation}

where parameters that fit the observed data will have a low cost and parameters that have a poor fit have a high cost. In the spectral modelling used in this study, the free parameters are the abundances of the endmembers ($w_i$ in \autoref{eq:linear-spectral-model}).

At each step in the MCMC chain, the walker selects a new set of parameter values to test $\vec{t}_{n}$ and moves to the new parameters with a probability proportional to the ratio of the cost functions $C(\vec{w}_n)/C(\vec{t}_{n})$. This means that if $\vec{t}_{n+1}$ is a better fit to the data, the walker will likely move to those parameters ($\vec{w}_{n+1} = \vec{t}_{n}$), otherwise it will likely remain in the same place ($\vec{w}_{n+1} = \vec{w}_{n}$). Therefore, each walker generally moves towards and then remains in a region of parameter space which has a good fit to the data.

After an initial `burn-in' period, the probability of the walker occupying a given set of parameter values is proportional to the posterior probability of the set of parameter values being consistent with the observed data. Therefore, the positions of a large ensemble of walkers can be sampled to produce a posterior probability distribution of the parameter values.

When applied to spectral modelling, MCMC enables the simulation of reflectance spectra, and the calculation of the posterior probability distribution of endmembers abundances being consistent with an observed spectrum. The posterior distribution for each endmember can be sampled to calculate the most likely abundance value and its associated uncertainty.

The use of MCMC techniques allows more robust modelling of reflectance spectra than simple linear optimization by quantifying the uncertainties of and correlations between endmember abundances \cite{lapotre2017probabilistic}. This ultimately allows us to measure the confidence of different detections, which is especially important for observations with relatively low spectral resolutions and potentially degenerate fit results where multiple different compositional mixes are possible \cite{brown2013salts}.

Our MCMC modelling uses the `emcee' library to directly calculate the walker Markov Chains. See \citet{foreman2013emcee} for a more detailed discussion of MCMC and the specific emcee implementation.

The only free parameters in our MCMC modelling are the abundances of each endmember ($w_i$ in \autoref{eq:linear-spectral-model}). All abundances are assumed to be equally likely, so a constant prior between 0 and 1 is used. This means that any potential solutions which would imply unphysical abundance values (i.e. less than 0\% or greater than 100\%) are rejected by the MCMC process.

We assume all possible combinations of abundance values are equally likely so we have a constant prior for all physically possible mixtures. Therefore, we have no prior bias between mixtures consisting of a single endmember and mixtures containing many endmembers. This is equivalent to the Dirichlet prior with all concentration parameters equal to one used in \citet{lapotre2017probabilistic}. The sum-to-one constraint, i.e. that the total abundance of all $N$ endmembers should be 100\% (\autoref{eq:sum-to-one-constraint}), is directly enforced by allowing the abundances of $N-1$ endmembers to vary then setting the final endmember's abundance to be $w_N=1-\sum_i^{N-1}w_i$. This is mathematically equivalent to rejecting all solutions where the abundances do not sum to one, but is computationally much more efficient.

Due to its underlying randomness, MCMC requires simulations using large numbers of walkers each with long chains to produce a stable output. This makes MCMC computationally intensive, typically requiring several minutes on a single processor core to run an MCMC fit for a single spectrum. Therefore, we used an optimised fitting routine to perform accurate fits across the whole observed disc within a reasonable timescale:

\begin{enumerate}
	\item Perform a simple linear fit to calculate optimized parameter values, $w_i^{\text{LF}}$.
	\item Generate initial positions, \verb|pos0|, for MCMC chains as a random Gaussian ball centred on $w_i^{\text{LF}}$ with a standard deviation of \verb|initialisation_sd|. These initial chain positions are corrected to ensure $0 \leq w_i \leq 1$ and $\sum_i{w_i} = 1$. Initialising around the linear fit values significantly decreases the required chain length, as the chain is initialised in a realistic region of the parameter space that is likely to be close to the region it will converge to.
	\item Run emcee for \verb|burn_in_steps| steps using \verb|pos0| as the initialisation. This `burns in' the chain to ensure the chain position is relatively independent of the initialisation.
	\item Continue running emcee in a series of runs with \verb|run_steps| steps. Each run is initialised with the final position of the previous run (i.e. it directly continues the chain). The abundance values are recorded for each successive run by calculating the median value over all walkers in the run.
	\item If the median values of \emph{every} parameter vary by less than \verb|convergence_difference| over \verb|convergence_runs| successive runs, the MCMC chain is assumed to have converged and the simulation is stopped. Otherwise, the simulation is automatically stopped at \verb|max_steps| steps to prevent indefinite calculations (though in practice this limit is never reached).
	\item The final abundance values and associated uncertainties are calculated from the posterior distribution of values produced by the final run. We take the median of the posterior distribution as the best estimate of the abundance and use the 16th and 84th percentiles of the distribution to calculate the 1-$\sigma$ uncertainty on this best estimate.
\end{enumerate}
This fitting routine has been tested over a wide range of inputs to ensure it remains accurate and the choice of different parameters do not influence the final calculated values. For example, initialisations with completely random values for \verb|pos0| ultimately converge to identical values as initialisations using linear fit values, but just take significantly longer to do so. Therefore, using the linear fit values is an effective shortcut to reduce calculation time without sacrificing accuracy. The values of the parameters used for our MCMC fitting routine are given in \autoref{tab:mcmc-parameters}.

The abundances of a single class of endmembers (e.g. all ice endmembers) can be combined into a single abundance distribution by summing the abundances of the individual endmembers at each time step for each walker. This produces a posterior distribution for the whole class of endmembers which often has a much smaller uncertainty than the individual endmember abundances that contribute to it. For example, if two specific endmembers have abundances of 10\% and 40\% respectively at one time step and then 30\% and 20\% at another time step, their individual uncertainties appear to be relatively large. However the uncertainty on the combination of these two endmembers is much smaller as at both time steps their summed abundance is 50\%.

\subsection{Spectral library}
\label{sec:spectral-library}
\begin{deluxetable*}{lll}
	\tablecaption{Spectral library \label{tab:spectral-library}}
	\tablehead{\colhead{Endmember} & \colhead{Name} & \colhead{Reference}}
	\startdata
	\ce{H2O} (refractive indices) & Water ice & \citet{grundy1998temperature} \\
	\hline
	\ce{H2SO4.6.5H2O}  & Sulphuric acid & \citet{carlson1999sulfuric} \\
	\ce{H2SO4.8H2O} (\SI{5}{\micro m}) & & \citet{carlson1999sulfuric} \\
	\ce{H2SO4.8H2O} (\SI{50}{\micro m}) & & \citet{carlson1999sulfuric} \\
	\hline
	\ce{Mg(ClO3)2.6H2O}  & Magnesium chlorate & \citet{hanley2014reflectance} \\
	\hline
	\ce{Mg(ClO4)2.6H2O}  & Magnesium perchlorate & \citet{hanley2014reflectance} \\
	\hline
	\ce{MgCl2.2H2O}  & Magnesium chloride & \citet{hanley2014reflectance} \\
	\ce{MgCl2.4H2O}  & & \citet{hanley2014reflectance} \\
	\ce{MgCl2.6H2O}  & & \citet{hanley2014reflectance} \\
	\hline
	\ce{MgSO4.6H2O}  & Magnesium sulphate & \citet{dalton2012low} \\
	\ce{MgSO4.7H2O}  & & \citet{dalton2012low} \\
	\ce{MgSO4 Brine}  & & \citet{dalton2007linear} \\
	\hline
	\ce{Na2SO4.10H2O}  & Mirabilite & \citet{dalton2007linear} \\
	\hline
	\ce{Na2Mg(SO4)2.4H2O}  & Sodium magnesium sulphate & \citet{dalton2012low} \\
	\hline
	\ce{NaClO4}  & Sodium perchlorate & \citet{hanley2014reflectance} \\
	\ce{NaClO4.2H2O}  & & \citet{hanley2014reflectance} \\
	\hline
	\ce{NaCl}  & Sodium chloride & \citet{hanley2014reflectance} \\
	\enddata
\end{deluxetable*}
\begin{figure*}
	\centering
	\includegraphics[width=0.9\linewidth]{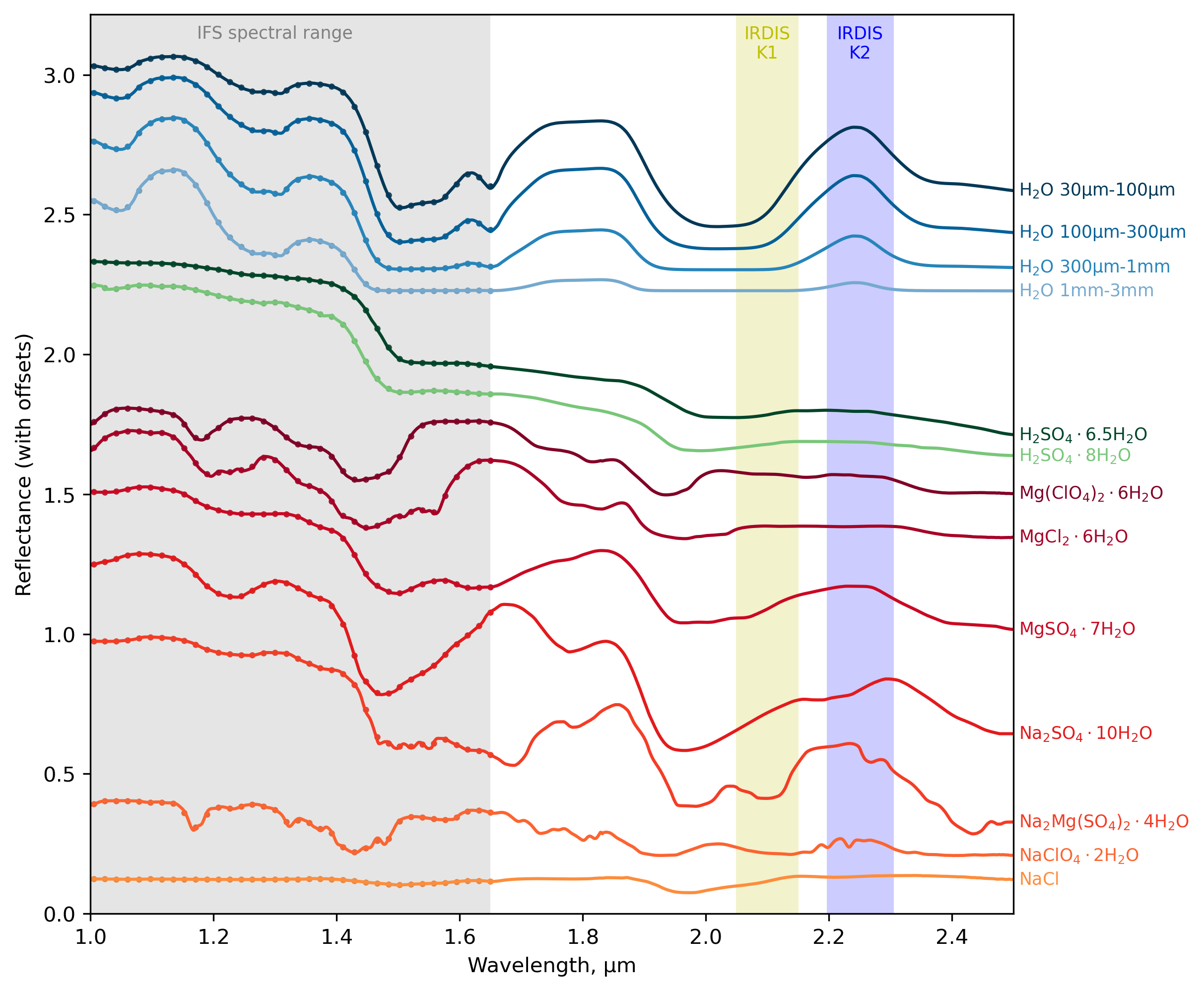}
	\caption{Example endmembers used for spectral modelling. The shaded areas show the spectral range covered by the SPHERE IFS instrument and the two IRDIS filters, and the dots show the reflectance in the IFS spectral bins. The NIMS spectra used in this study cover the whole spectral range shown in this figure.
		\label{fig:example-endmembers}}
\end{figure*}

The full list of endmembers in our spectral library are given in \autoref{tab:spectral-library} and selected spectra are shown in \autoref{fig:example-endmembers}. These cryogenic reference spectra include hydrated sulphuric acid \cite{carlson1999sulfuric} and a variety of hydrated salts \cite{hanley2014reflectance, dalton2012low, dalton2007linear}. The sulphuric acid spectra \cite{carlson1999sulfuric} do not cover wavelengths below \SI{1}{\micro\m}, so this limits our modelling to cover the \SIrange{1}{2.5}{\micro\m} spectral range. Our spectral library includes all available laboratory spectra covering the SPHERE wavelength range measured in Europa-like conditions and particular care was taken to ensure the inclusion of the various non-ice species detected in prior works \cite{carlson1999sulfuric,mccord2002brines,brown2013salts,ligier2016vlt,trumbo2019sodium}.

Our water ice reflectance spectra are calculated from measured refractive indices \cite{grundy1998temperature} using the Hapke bidirectional reflectance model \cite{hapke1993book}. Modelling was performed using a variety of simulated ice grain sizes ranging from \SI{1}{\micro m} to \SI{1}{cm} to fully explore the grain size parameter space. Initial models did not identify any extreme small (\SI{<30}{\micro m}) or large (\SI{>3}{mm}) grains, so our final model presented here covers grains ranging from \SI{30}{\micro m} to \SI{3}{mm}.

The final modelled water ice spectra use four size bins ranging from \SI{30}{\micro m} to \SI{3}{mm} where the spectrum is a blend of grain sizes within each bin. Each blended spectrum is calculated by modelling 250 discrete grain sizes linearly spaced within the size bin and then calculating the mean of these 250 spectra. This provides a more physically realistic simulation of the grain sizes, as we would not expect there to only be a single discrete ice grain size present on Europa's surface. This averaging also removes any residual Mie oscillations in the calculated spectrum that can occur when a single exact grain size is used for modelling \cite{hapke1993book}.

\section{Regions of interest}
\begin{figure*}
	\centering
	\includegraphics[width=\linewidth]{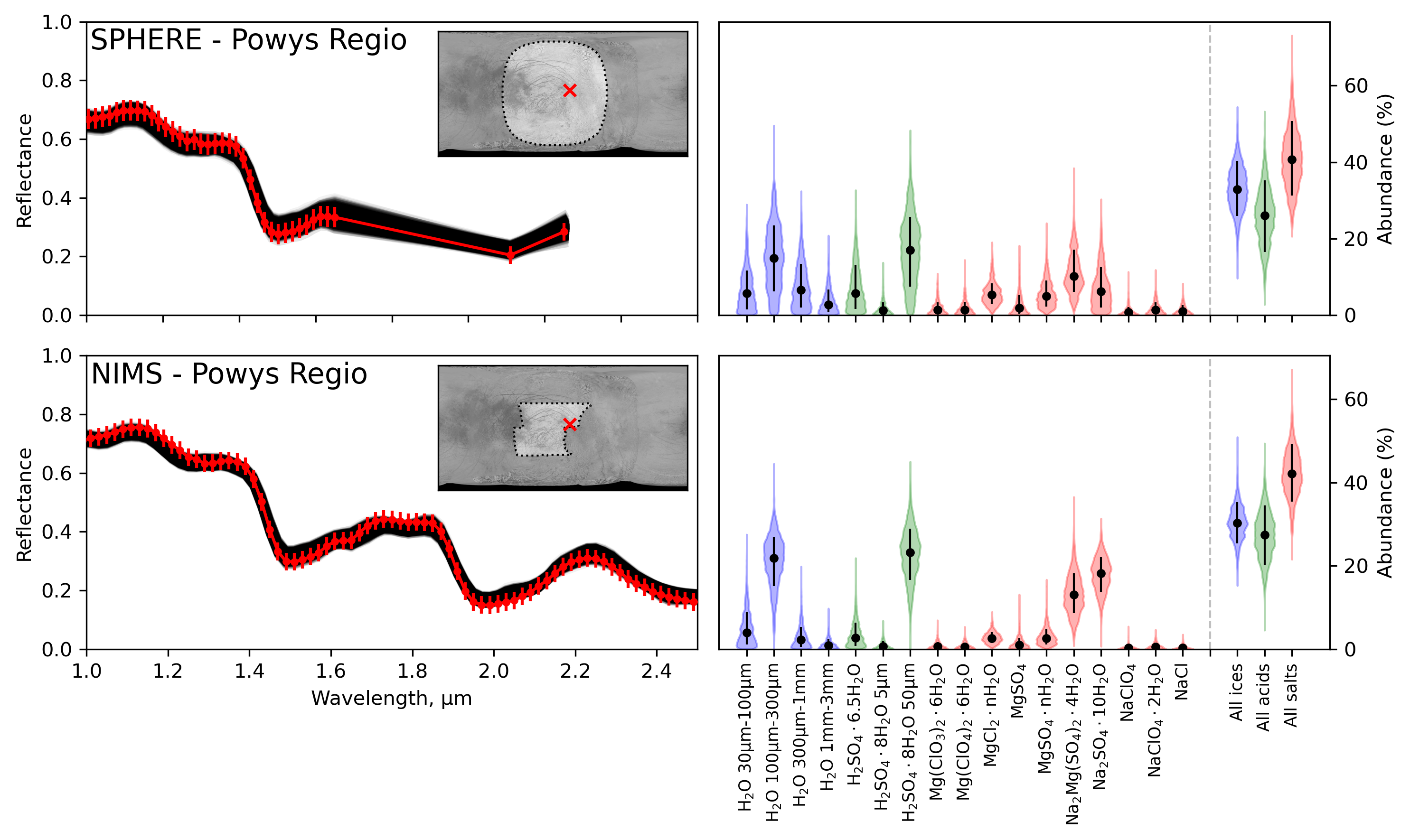}
	\includegraphics[width=\linewidth]{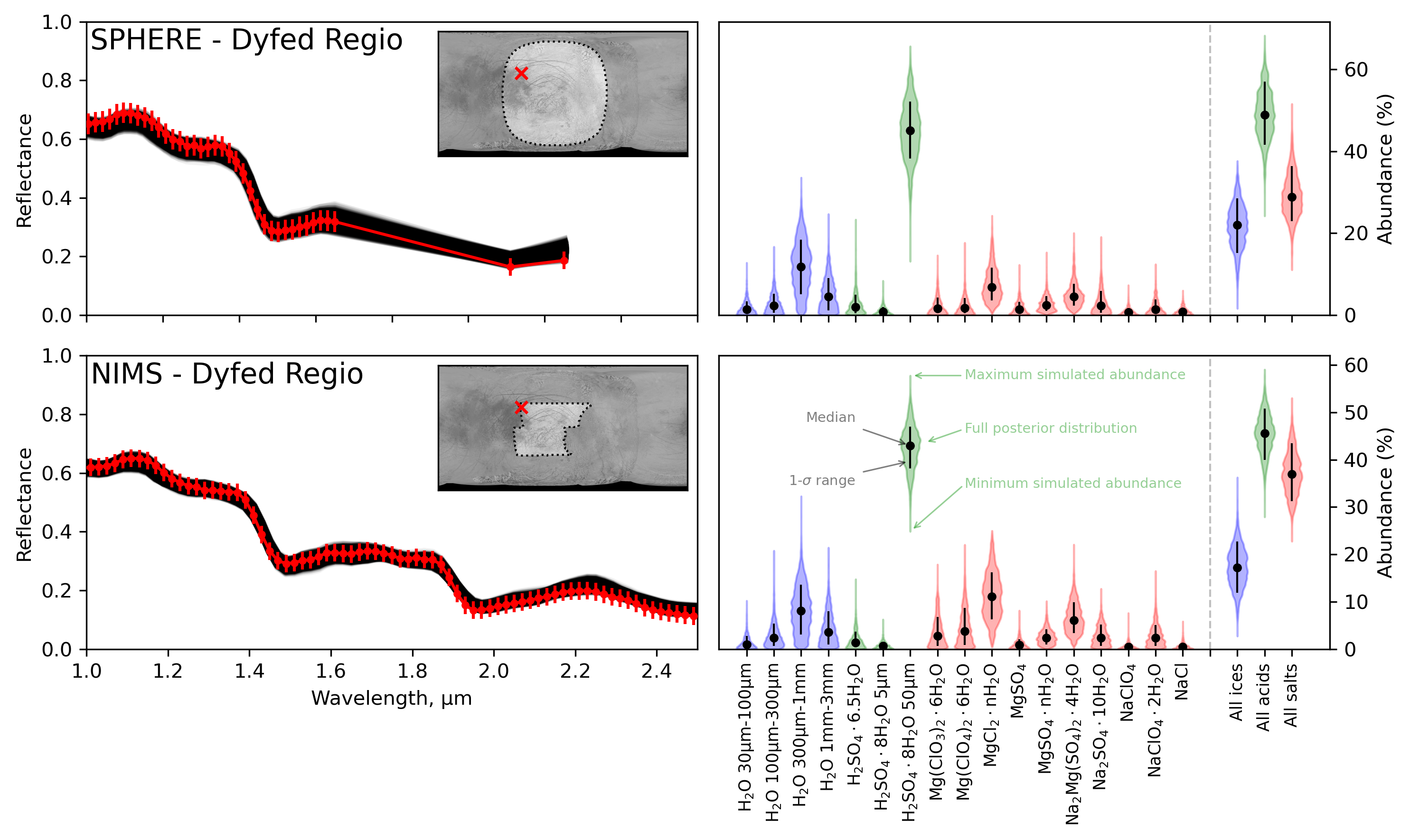}
	\caption{Fit result for spectra from Powys Regio and Dyfed Regio. The left hand column shows the observed spectrum (red) and the `family' of MCMC fitted spectra (black). The right hand column shows the violin plots for the fitted posterior abundance distributions where the shaded region shows the full posterior distribution of abundance values. The horizontal width of the shaded region shows the shape of the posterior distribution (where wider areas are more likely and narrower areas are less likely) and the vertical height of the shaded region shows the full range of simulated abundance values. The black dot shows the median, best-estimate, abundance and the black line gives the 1-$\sigma$ uncertainty around this estimate. Numerical values are provided in \autoref{tab:case-study-abundances}.
		\label{fig:violin-both}}
\end{figure*}
\begin{figure*}
	\centering
	\includegraphics[width=\linewidth]{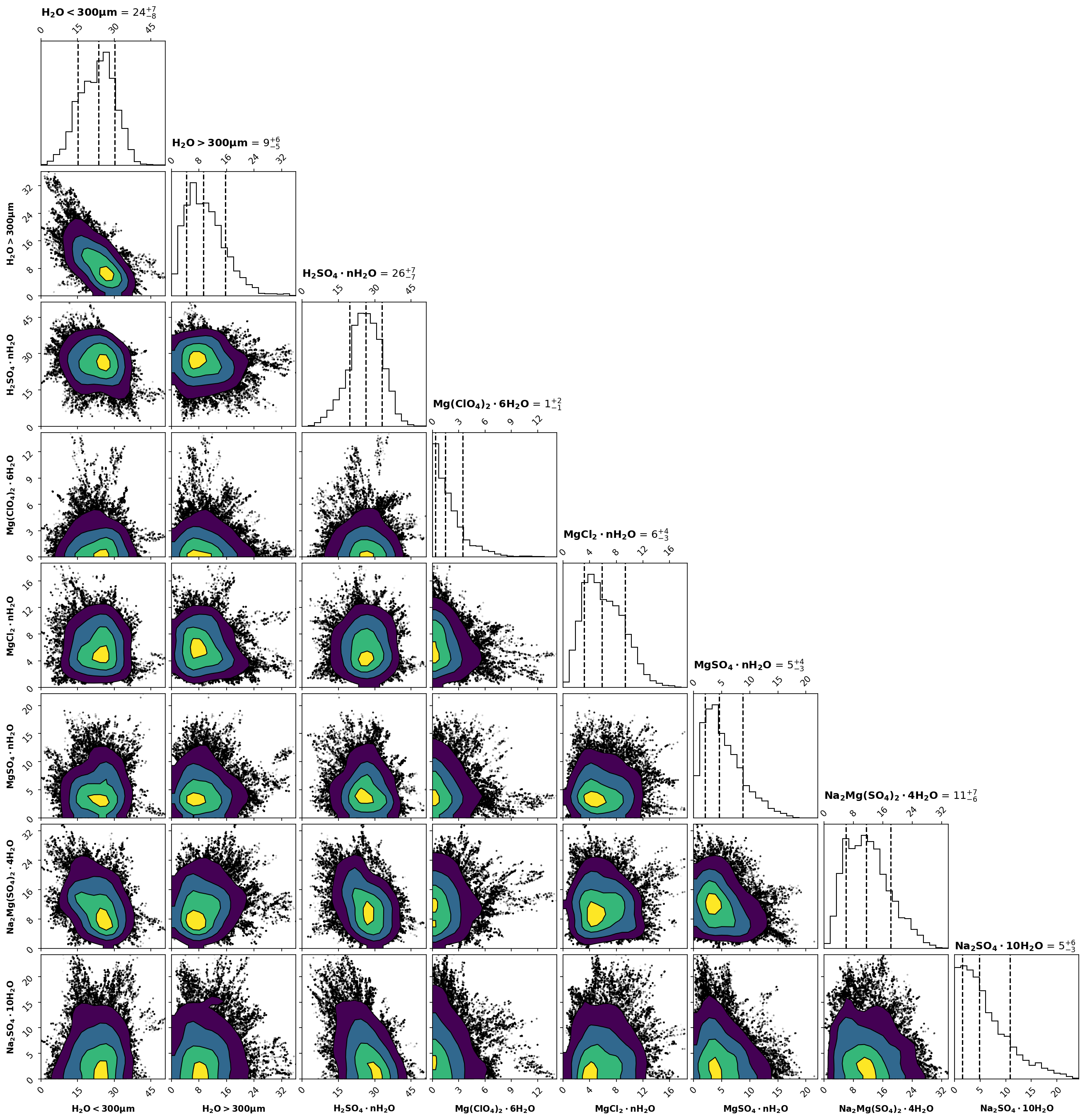}
	\caption{Corner plot showing relationship between endmember percentage abundances for the SPHERE fit in Powys Regio. The shaded 2D-histograms give the posterior abundance distribution with brighter colours indicating a higher density. In low density regions towards the edge of the distribution, individual points of the distribution are plotted. For clarity, some endmembers have been combined. Circular distributions imply no strong correlation between the abundance distributions of the two endmembers whereas skewed distributions imply a correlation between the abundances of those endmembers.
		\label{fig:corner-powys}}
\end{figure*}

Initial modelling was performed on regions of interest in Powys Regio (\coords{170}{5}) and Dyfed Regio (\coords{240}{30}). These locations both have high non ice content, as indicated by the weakness of the \SI{1.5}{\micro m} water ice absorption in \autoref{fig:1.5-band-map}. These also sample both the leading (Powys Regio) and trailing (Dyfed Regio) hemispheres, so provide a useful comparison between the non-ice material in the two hemispheres and their associated plasma and radiation environments. In addition to comparing the composition of the two locations, we can compare the SPHERE and NIMS fits for the same location to understand the limitations of the datasets.

The MCMC fit results for both locations are summarised in \autoref{fig:violin-both}, where the violin plots on the right hand side show the fitted abundance distributions for each endmember and family of endmembers. Both the SPHERE and NIMS datasets show consistent results, with a roughly even mixture of ices, acids and salts. As expected, Dyfed Regio on the trailing hemisphere has a higher acid abundance, and there are differences in the salt mixtures for each location (discussed in more detail in \autoref{sec:salt-results}).

It is notable that the uncertainty on the fitted abundances is larger for some endmembers in the SPHERE dataset. For example SPHERE is less able to discriminate between different water ice grain sizes in Powys Regio, and is less able to rule out the detections of individual salts. This demonstrates the utility of the additional spectral range covered by NIMS in helping to lift degeneracies between different endmembers and increase the confidence of their detections. However, the best estimate values are consistent for both datasets, suggesting that the missing spectral range for SPHERE is not leading to any spurious results.

The MCMC results can also be summarised using corner plots that show the relationship between posterior distributions of pairs of endmembers from the modelling of an individual spectrum. \autoref{fig:corner-powys} shows the corner plot result for the SPHERE Powys Regio fit. Generally, most of the 2D-histograms in \autoref{fig:corner-powys} are roughly circular, implying relatively little relationship between the individual endmember abundances distributions. However some of the 2D-histograms, particularly between water ice grains (top left), are clearly skewed, implying correlation between the endmember abundance distributions. For the case of small and large water ice grains, this negative covariance implies that much of the uncertainty in individual endmember abundance is due to the slight degeneracy between grain sizes. This can be seen in the violin plots in \autoref{fig:violin-both} where there is relatively large uncertainty on some individual ice grain size endmembers, but these cancel out, leaving a smaller uncertainty on the total water ice abundance.

\section{Compositional maps}
MCMC fitting was performed for each observed spectrum for both the SPHERE and NIMS datasets. This produces posterior abundance distributions for each location that were sampled to create maps of best estimate abundances for different species, showing their spatial distributions.

\subsection{Hydrated sulphuric acid}
\begin{figure*}
	\centering
	\includegraphics[width=0.75\linewidth]{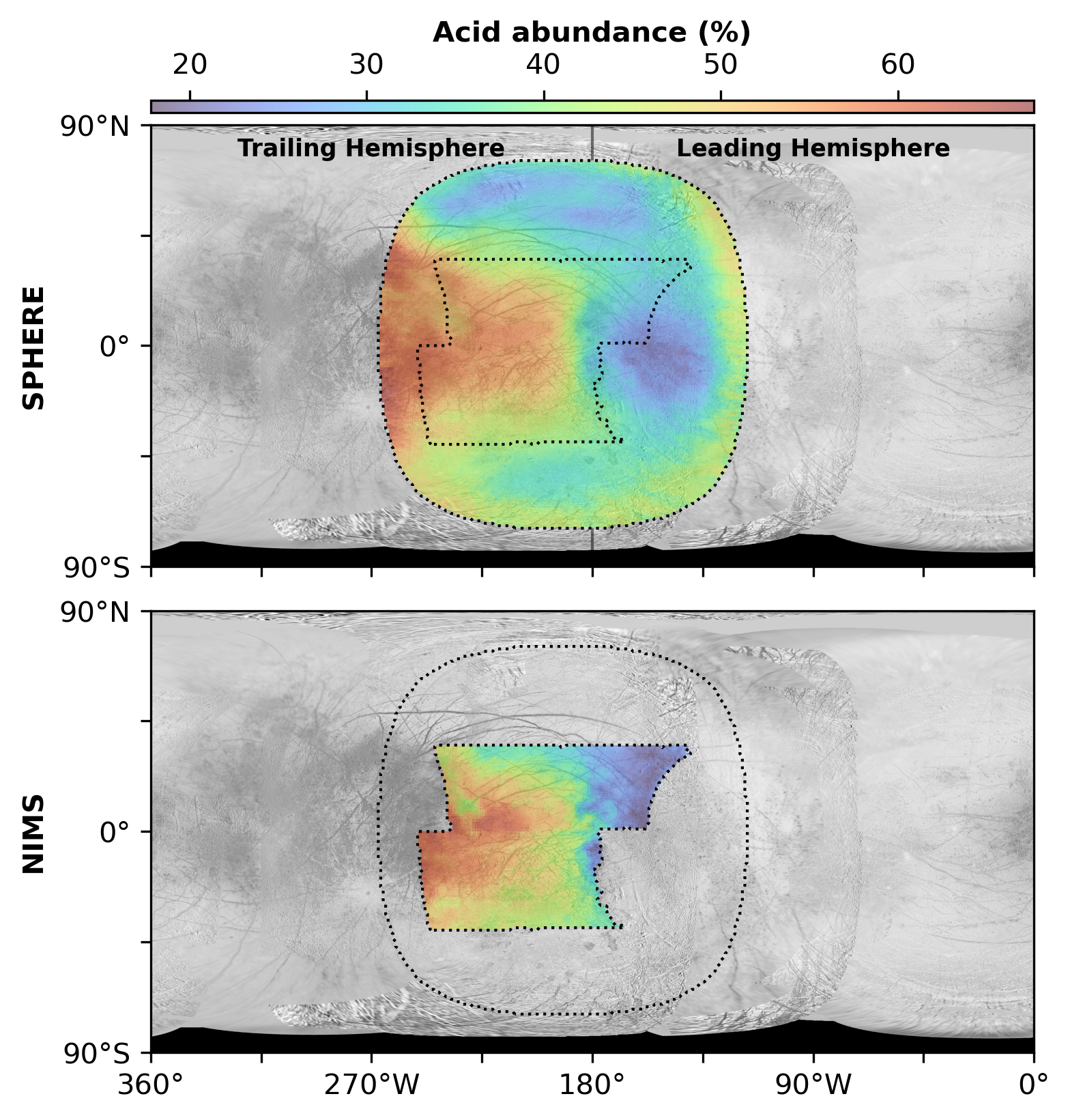}
	\caption{Hydrated sulphuric acid spatial distribution for the SPHERE and NIMS datasets. The values shown here are the best estimates that are calculated as the median of the posterior abundance distribution for each location. The acid abundance is highest towards the trailing apex (\coords{270}{0}) and lower towards the leading hemisphere and high latitudes. The slight increase in acid abundance towards the edge of the SPHERE dataset is likely to be caused by residual errors from the photometric correction. The typical uncertainty on the abundance values is $\pm10$ percentage points for SPHERE and $\pm7$ percentage points for NIMS.
		\label{fig:acid-map}}
\end{figure*}
The hydrated sulphuric acid distribution shown in \autoref{fig:acid-map} is mainly concentrated towards the trailing hemisphere, with lower abundances towards the leading hemisphere and at higher latitudes. There is no strong correlation with geological units, and the distribution for both SPHERE and NIMS follows the expected bullseye distribution centred on the trailing apex with a clear distinction between the leading and trailing hemispheres. This spatial distribution is consistent with exogenic plasma bombardment being the dominant source of the sulphuric acid hydrate present on Europa's surface.

The peak abundance is 68\% for SPHERE (1-$\sigma$ range 60\% to 75\%) and 65\% for NIMS (1-$\sigma$ range 60\% to 70\%), both of which occur around the trailing apex. These values are similar to the $\sim65\%$ maximum abundance in \citet{ligier2016vlt} though lower than the $\sim90\%$ reported in \citet{carlson2005distribution}. However it is important to note that these previous studies have full spatial coverage of the trailing apex whereas the SPHERE and especially NIMS datasets shown in \autoref{fig:acid-map} only cover some of the eastern part of the trailing hemisphere, so the values are not directly comparable.

\subsection{Water ice}
\begin{figure*}
	\centering
	\includegraphics[width=0.75\linewidth]{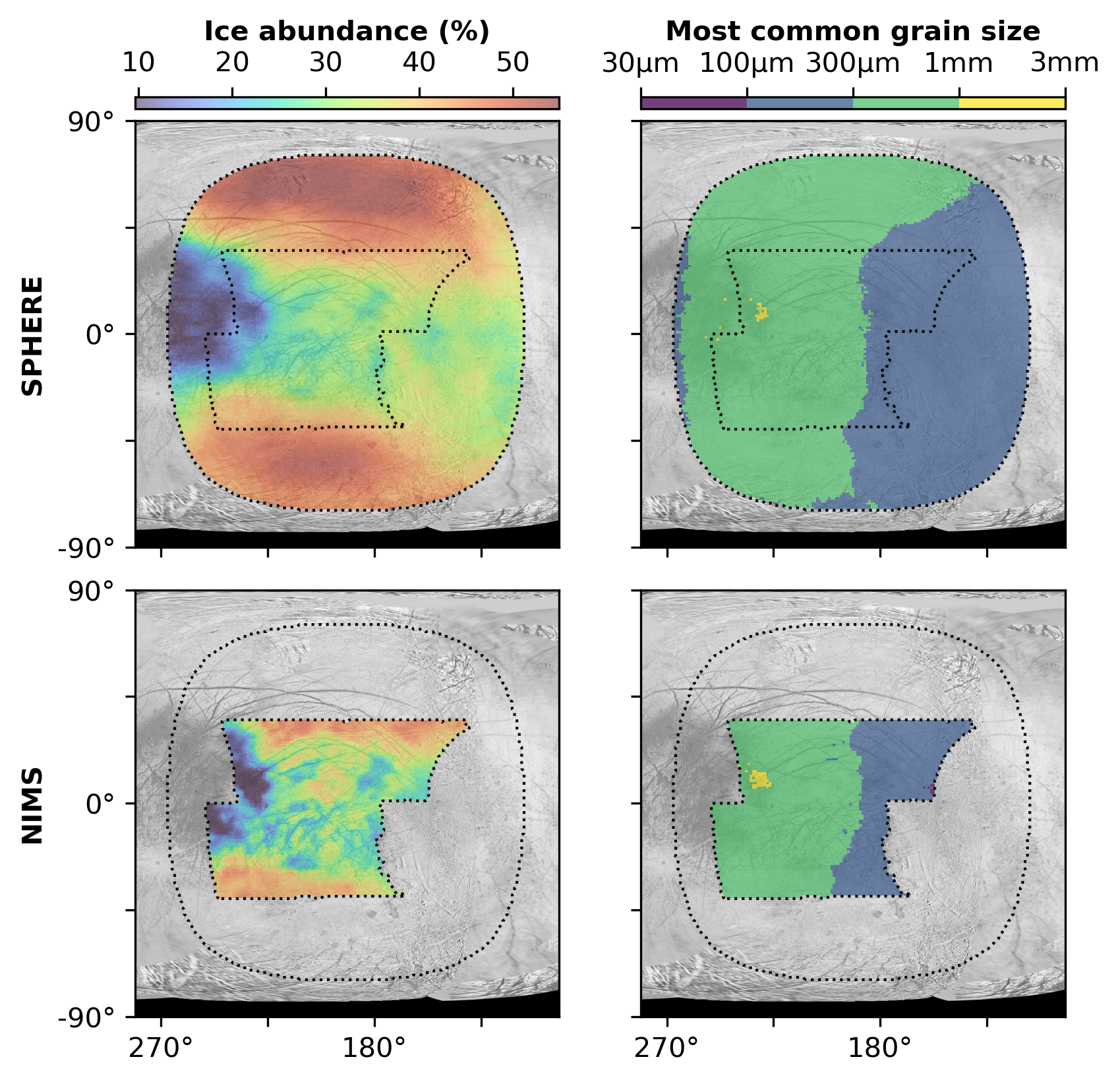}
	\caption{Water ice spatial distributions for SPHERE (top) and NIMS (bottom). The total water ice abundance (left) is lowest towards the trailing apex and highest at high latitudes where abundances are $>50\%$. The grain size column (right) shows the ice size endmember with the highest individual abundance. The ice grain size varies with longitude, with larger (\SI{300}{\micro\m} to \SI{1}{mm}) grains more abundant on the trailing hemisphere and smaller grains (\SI{100}{\micro\m} to \SI{300}{\micro\m}) more abundant on the leading hemisphere.
		\label{fig:ice-map}}
\end{figure*}
The modelled water ice distribution in \autoref{fig:ice-map} shows the same spatial pattern as the \SI{1.5}{\micro m} (\autoref{fig:1.5-band-map}) and \SI{2}{\micro m} (\autoref{fig:irdis-two_color}) absorption bands with high abundance at high latitudes and lower abundance towards the trailing apex. The water ice distribution is highly anti-correlated with the sulphuric acid distribution, with the combined hydrated sulphuric acid and ice abundance $\sim80\%$ when averaged over the observed area. This anti-correlation is a real surface feature (and not a spectral degeneracy), where the spatial distribution of water ice abundance is driven by the absence of contaminants, mainly acids in the trailing hemisphere and to a lesser extent salts in the leading hemisphere. The highest water ice abundances occur at high latitudes, where the exogenic plasma bombardment intensity is significantly lower.

As found in previous studies \cite{dalton2012europa, ligier2016vlt}, the majority of the surface is dominated by \SI{100}{\micro m} to \SI{1}{mm} grains, with larger grains (\SI{>300}{\micro m}) on the trailing hemisphere and northern latitudes and smaller grains on the leading hemisphere. There also appears to potentially be a small localised area of \SI{>1}{mm} grains in Dyfed Regio, however the overall ice abundance in this region is very low, so the confidence of this detection of larger grains is lower.

\subsection{Salts}
\label{sec:salt-results}
\begin{figure*}
	\centering
	\includegraphics[width=\linewidth]{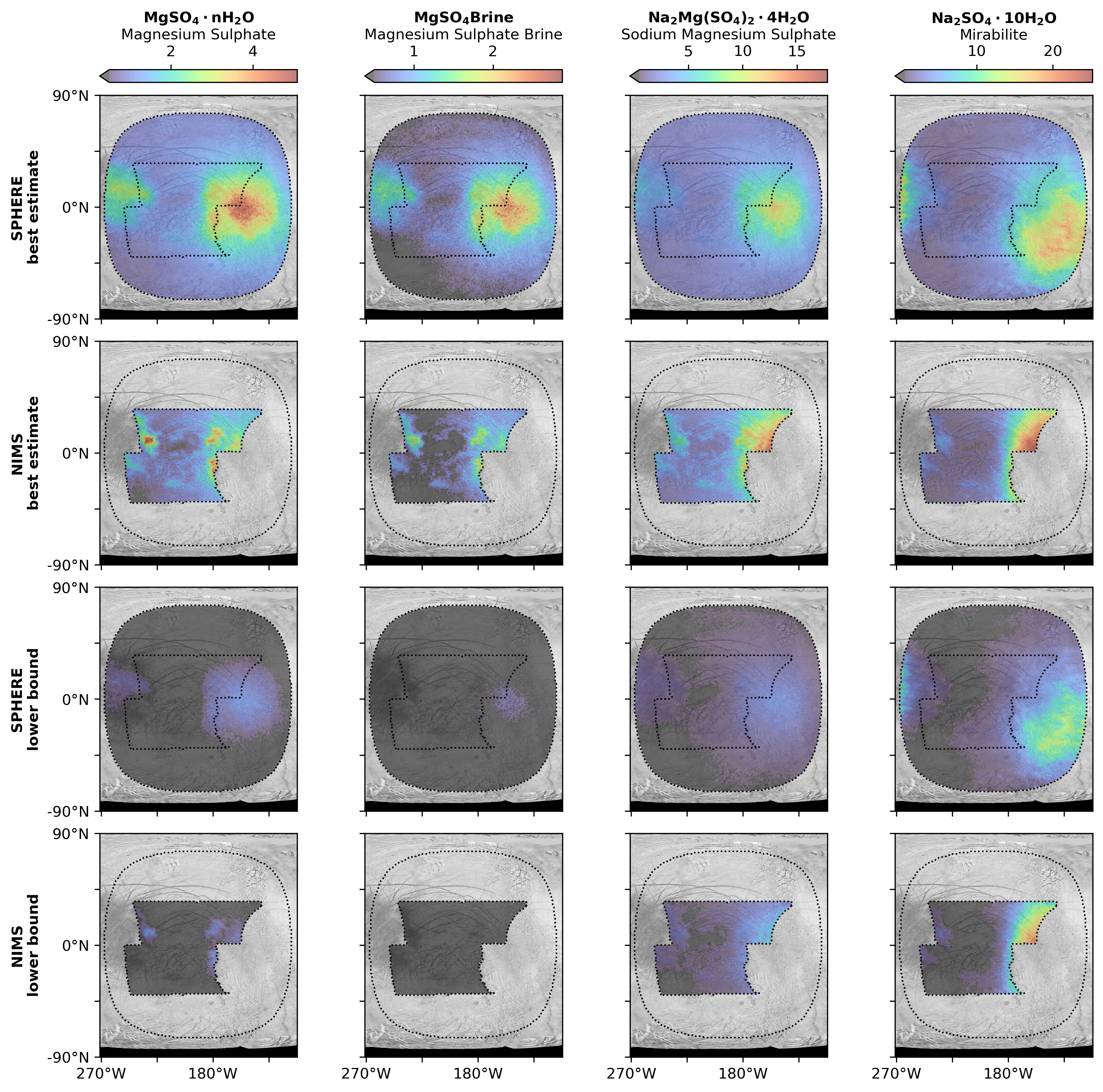}
	\caption{Spatial distribution of sulphate salt abundances. The top rows show best estimate abundances (median of posterior distribution) and the bottom rows show the 1-$\sigma$ lower bound of the posterior abundance distributions. Lower bounds close to zero suggest that the specific endmember cannot be confidently detected. Abundance values less then 0.5\% are shown in black. Note the different abundance scale for each salt species.
		\label{fig:salt-map-s}}
\end{figure*}

A variety of hydrated salts were identified in contaminated regions, especially around Dyfed Regio and Powys Regio. As with previous analyses of NIMS observations \cite{mccord2002brines, dalton2007linear, shirley2010europa}, sulphates appear to be the main salts present on Europa's surface (see \autoref{fig:salt-map-s}). Magnesium bearing salts (\ce{MgSO4.nH2O}, \ce{MgSO4} brine and \ce{Na2Mg(SO4)2.4H2O}) appear correlated with dark terrain, particularly Powys Regio. As shown in the lower panels of \autoref{fig:salt-map-s}, the lower bounds on the abundances of magnesium sulphates are all very low and close to zero. This suggests that the uncertainties and degeneracies between the salt spectra mean it is not possible to positively identify any individual magnesium sulphate salts with the SPHERE and NIMS spectral resolutions.

Mirabilite (\ce{Na2SO4.10H2O}) appears more abundant in Eastern areas of the observations in \autoref{fig:salt-map-s}, with a relatively clear distinction between low abundances in the trailing hemisphere and higher abundances into the leading hemisphere. Mirabilite has the highest abundances ($\sim20\%$) of any of the modelled salts, and the corresponding non-zero lower bounds suggest that it is likely to be present on the leading hemisphere. The modelled abundances suggest an upper bound of $\sim30\%$ for mirabilite abundance on the anti-jovian hemisphere.

\begin{figure*}
	\centering
	\includegraphics[width=\linewidth]{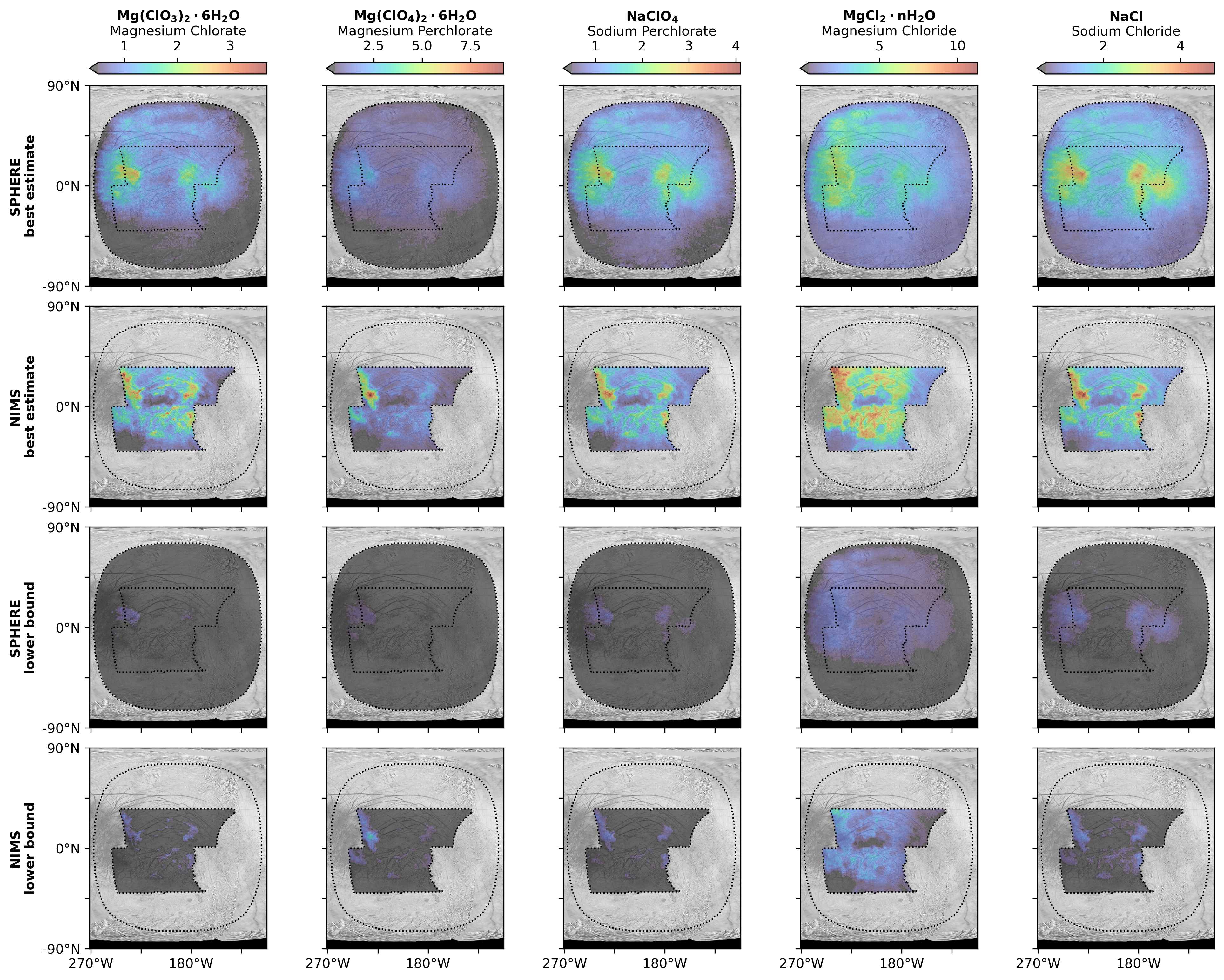}
	\caption{Spatial distribution of chlorine bearing salt abundances. As with \autoref{fig:salt-map-s}, the top rows show the best estimate abundances and the bottom rows show 1-$\sigma$ lower bounds. Abundance values less then 0.5\% are shown in black. Note the different abundance scale for each salt species.
		\label{fig:salt-map-cl}}
\end{figure*}

Chlorinated salts (\autoref{fig:salt-map-cl}) generally have lower best estimate abundances than the sulphate salts and all have lower bounds close to zero, implying it is difficult to positively identify individual salts, particularly with SPHERE. The spatial distribution of the chlorinated salts appears more globally uniform than the sulphate salts with slightly higher abundances around geological units.

Hydrated magnesium chlorate (\ce{Mg(ClO3)2.6H2O}), perchlorate (\ce{Mg(ClO4)2.6H2O}) and sodium perchlorate (\ce{NaClO4}) all have very low abundances and appear slightly correlated with geological units. Magnesium perchlorate has a single region of relatively high abundance in the NIMS observation of Dyfed Regio which is consistent with the location of the highest abundance found in \citet{ligier2016vlt}. Sodium chloride (\ce{NaCl}) and sodium perchlorate also have a low abundance in our observations, with Powys Regio and Dyfed Regio the only areas with tentative detections. UV absorptions attributed to sodium chloride have been previously detected around Europa's leading apex using HST \cite{trumbo2019sodium}, however this region is outside of our observed area, so is consistent with our low abundances.

Hydrated magnesium chloride (\ce{MgCl2.nH2O}) appears the most abundant chlorinated salt, with best estimate abundances up to 10\% in the NIMS dataset. The distribution of magnesium chloride appears correlated with disrupted geological units, particularly Dyfed Regio and the northern lineae in the SPHERE observation and Argadnel Regio in the NIMS observation. This distribution appears consistent with \citet{ligier2016vlt} who found abundances of $\sim 10\%$ around Dyfed Regio.

\section{Discussion}
Compositional contrasts on Europa's surface appear to be primarily driven by exogenic plasma bombardment and the resulting longitudinal gradients in rates of radiation driven processes. As with previous studies \cite{carlson2005distribution, dalton2012europa, ligier2016vlt}, the leading-trailing contrast in acid abundance strongly suggests an exogenic origin for the sulphuric acid as the acid abundances appear correlated with sulphur ion bombardment.

Similarly, the leading-trailing hemisphere dichotomy of ice grain sizes suggests that exogenic processes are likely to be responsible for the contrast in grain sizes. As described in \citet{clark1983frost}, more intense radiation on the trailing hemisphere will lead to a faster sputtering rate for water ice grains compared to the leading hemisphere. This sputtering erodes grains at a rate proportional to the surface area of the grain, preferentially destroying smaller grains due to their higher surface area to volume ratio. Therefore, the leading hemisphere's lower sputtering rate will increase the lifetime of smaller grains, thus reducing the average grain size on the leading hemisphere \cite{cassidy2013magnetospheric}.

In addition to the longitudinal variation in water ice grain size, the SPHERE data also show some latitudinal contrast, with larger grains extending further eastwards in the northern hemisphere. The cause of this latitudinal variation is unclear, especially given that it is not repeated in the southern hemisphere, suggesting that thermal effects are unlikely to be the cause as these would be expected to be symmetric around the equator. Therefore, this region of larger ice grains may be caused by some endogenic process, potentially related to the age of the surface.

Cratered areas and ridged plains, particularly at high latitudes, show high water ice abundance. Chaos areas and bands on the other hand appear more contaminated, especially Dyfed Regio which is located near the trailing apex and therefore receives the most intense exogenic plasma bombardment. Best estimate salt abundances are highest in the chaos areas in Dyfed and Powys Regio.

The uncertainties on the modelled salt abundances demonstrate how the low spectral resolution SPHERE and NIMS data are consistent with a variety of different salt mixtures. The lack of unique distinguishing spectral features at these low spectral resolutions means that the endmember spectra are partially degenerate, meaning that the observed spectra generally cannot be used to positively identify individual salt species (see \autoref{fig:salt-map-s} and \autoref{fig:salt-map-cl}). This makes detailed analysis of the salt distributions and their potential origins difficult due to the wide variety of potential mixtures. However, the high spatial resolution offered by SPHERE and NIMS can be used to infer the likely spatial distribution the salts would have if they are present. These spatial distributions can then be used in comparison with high spectral resolution observations that have more confidently identified specific species in a general area.

The clearest trend in our data was the higher mirabilite (\ce{Na2SO4.10H2O}) abundance in the leading hemisphere (highest abundance at $\sim\ang{130}$W) contrasting with the higher magnesium sulphate abundances towards and into the trailing hemisphere (highest abundance at $\sim\ang{160}$W). This is very similar to the longitudinal contrasts in hexahydrite (\ce{MgSO4.6H2O}) and mirabilite abundances found in \citet{shirley2010europa}. \citet{mccord2001thermal} found that magnesium sulphates are more stable than sodium sulphates against thermal dehydration and radiolysis, meaning that higher radiation on the trailing hemisphere will destroy mirabilite at a faster rate. Therefore, as with acids and ices, the exogenic radiation environment may explain these observed longitudinal compositional contrasts.

Magnesium ions have not been detected in the Io plasma torus \cite{carlson2009europa}, so the most plausible source for \ce{Mg} ions is Europa's sub-surface ocean. \citet{mccord2002brines} suggested that magnesium sulphates may originate directly from the sub-surface ocean whereas \citet{brown2013salts} suggested that magnesium sulphate is a radiation product of exogenic sulphur ions and magnesium salts already present on Europa's surface. Our observed magnesium sulphate distributions (\autoref{fig:salt-map-s}) appear broadly consistent with both hypotheses, with magnesium sulphates appearing mainly in geological units in regions where \citet{brown2013salts} detected magnesium sulphate. SPHERE observations with increased longitude coverage would allow more detailed investigation of the radiation product hypothesis, although the detection of a distinctive magnesium sulphate absorption feature mainly on the trailing hemisphere in high spectral resolution observations by \citet{brown2013salts} appears compelling. This \SI{2.07}{\micro\m} absorption feature is very narrow, so is not detectable with NIMS' spectral resolution.

Chlorine bearing salts, particularly magnesium chloride, appear correlated with geological units such as lineae and darker areas (\autoref{fig:salt-map-cl}). The magnesium chloride abundances are low ($\lesssim 10\%$), but appear consistent with the abundances and spatial distributions found in \citet{ligier2016vlt}. The strong correlation with geological units suggests an endogenic origin for these salts. If chlorine ions are present in Europa's sub-surface ocean, magnesium chloride would form directly in the ocean. Therefore, this observed magnesium chloride may have originated directly from the ocean \cite{ligier2016vlt} and would provide the magnesium source for any radiation produced magnesium sulphate \cite{brown2013salts}.

Magnesium perchlorate appears correlated with both the magnesium chloride distribution and the exogenic plasma bombardment centred on the trailing apex (\autoref{fig:salt-map-cl}). This is consistent with the endogenic magnesium chloride being converted to magnesium perchlorate under the high radiation ion bombardment in the trailing hemisphere around Dyfed Regio \cite{carrier2015origins,ligier2016vlt}.

As demonstrated by the uncertainties in salt abundances produced by the MCMC fitting routine, broader wavelength coverage and higher spectral resolution data will be necessary to positively identify or rule out the presence of many individual salts on Europa's surface. Future Juno observations of Europa will be able to use JIRAM's high spectral resolution and the very high spatial resolution offered by Juno's position in the Jupiter system to study and constrain Europa's non-ice surface composition at small scales \cite{mishra2021bayesian}. The next generation of instruments such as ELT/HARMONI and JWST/NIRSPEC will enable spectroscopy to identify narrow features, such as the \SI{2.07}{\micro m} magnesium sulphate absorption in \citet{brown2013salts}, whilst also acquiring high spatial resolution mapping to identify small-scale compositional contrasts.

\section{Conclusion}
\label{sec:conclusion}
The near-infrared spectral observations with SPHERE and NIMS have enabled compositional mapping of Europa's anti-jovian hemisphere. The two instruments' datasets appear generally consistent, demonstrating the utility of SPHERE's low spectral resolution and high spatial resolution ground-based spectroscopy for solar system targets.

Our reduction pipeline has enabled the use of SPHERE to observe extended targets by removing the banding pattern likely caused by crosstalk between different lenslets. The use of the Oren-Nayar reflectance model \cite{oren1994generalization}, combined with a diffraction model, allows photometric correction of observations to emission angles $>\ang{70}$. This allows mapping to extend to higher latitudes than previous studies, observing $>90\%$ of the observed disc. The limiting factor is generally now the low spatial resolution caused by the extreme viewing angles at the edge of the disc.

The use of MCMC modelling for spectral fits enables detailed uncertainty estimation on final fitted abundances. This is particularly useful for the relatively low-abundance hydrated salts, where the MCMC uncertainties quantitatively show that confident detection of individual salt endmembers is generally not possible with these datasets due to the significant degeneracies between hydrated salt spectra. High spatial resolution observations, such as with SPHERE, can be used to complement other observations by spatially mapping species which have been more confidently detected with higher spectral resolution, lower spatial resolution observations. Future studies may use further statistical techniques such as the Beysian information criterion \cite{schwarz1978estimating} to help to select between different potential models of Europa's surface composition.

The identified spatial distributions of species appears consistent with previous studies of Europa, with high acid abundance and larger ice grains on the trailing hemisphere and a variety of potential hydrated salts. Future studies will help to further constrain Europa's surface composition by expanding the spatial and spectral coverage and resolution of near-infrared observations. Higher spectral resolution observations (e.g. JWST and ELT/HARMONI) and laboratory reference spectra measurements will help to constrain salt abundances by identifying narrow characteristic features in the spectra, while high spatial resolution spacecraft observations will enable accurate geolocation of compositional features.

\acknowledgments
Based on observations collected at the European Organisation for Astronomical Research in the Southern Hemisphere under ESO programme 60.A-9372(A). Oliver King was supported by a Royal Society studentship grant at the University of Leicester. Leigh Fletcher was supported by a Royal Society Research Fellowship and European Research Council Consolidator Grant (under the European Union’s Horizon 2020 research and innovation programme, grant agreement No 723890) at the University of Leicester. This research used the ALICE High Performance Computing Facility at the University of Leicester. We are extremely grateful to Fraser Clarke and Niranjan Thatte for their support acquiring and reducing the data.

\appendix
The datasets used in this study can be found at \url{https://doi.org/10.5281/zenodo.6034735}.

\begin{deluxetable}{lrl}[h]
	\tablecaption{Parameters used for our MCMC fitting routine. \label{tab:mcmc-parameters}}
	\tablehead{\colhead{Name} & \colhead{Value} & \colhead{Description}}
	\startdata
	\tt nwalkers & 300 & Number of walkers used for each simulation\\
	\tt initialisation\_sd & 0.01  & Standard deviation of random gaussian ball used to initialise the walker positions\\
	\tt convergence\_difference & 0.02 & Threshold for maximum allowed variation to determine convergence\\
	\tt convergence\_runs & 5 & Number of runs required for convergence check\\
	\tt burn\_in\_steps & 5000 & Number of steps to `burn in' chain after it is initialised\\
	\tt run\_steps & 1000 & Number of steps in each run after burn in\\
	\tt max\_steps & 500000 & Maximum number of steps allowed\\
	\enddata
\end{deluxetable}

\begin{deluxetable}{lrrrr}[h]
	\tablecaption{Abundances values for case studies shown in \autoref{fig:violin-both}. \label{tab:case-study-abundances}}
	\tablehead{
		\colhead{} & \multicolumn{2}{c}{Powys Regio} & \multicolumn{2}{c}{Dyfed Regio}\\
		\colhead{Endmember} & \colhead{SPHERE} & \colhead{NIMS} & \colhead{SPHERE} & \colhead{NIMS}
	}
	\startdata
	\ce{H2O} \SI{30}{\micro\m}-\SI{100}{\micro\m} & $5.6^{+7.9}_{-4.1}\%$ & $4.1^{+4.8}_{-2.9}\%$ & $1.3^{+2.0}_{-1.0}\%$ & $1.2^{+1.4}_{-0.8}\%$\\
	\ce{H2O} \SI{100}{\micro\m}-\SI{300}{\micro\m} & $15.7^{+8.1}_{-8.8}\%$ & $21.7^{+5.7}_{-5.7}\%$ & $2.9^{+3.7}_{-2.2}\%$ & $2.6^{+3.3}_{-2.0}\%$\\
	\ce{H2O} \SI{300}{\micro\m}-\SI{1}{mm} & $5.8^{+6.4}_{-4.1}\%$ & $2.2^{+2.9}_{-1.6}\%$ & $12.5^{+6.3}_{-7.3}\%$ & $7.6^{+5.5}_{-4.8}\%$\\
	\ce{H2O} \SI{1}{mm}-\SI{3}{mm} & $2.7^{+3.8}_{-2.1}\%$ & $0.7^{+1.4}_{-0.5}\%$ & $3.6^{+4.0}_{-2.7}\%$ & $3.9^{+4.5}_{-2.9}\%$\\
	\hline
	\ce{H2SO4.6.5H2O} & $5.9^{+6.4}_{-4.3}\%$ & $2.2^{+3.2}_{-1.6}\%$ & $1.9^{+3.0}_{-1.5}\%$ & $1.5^{+2.2}_{-1.1}\%$\\
	\ce{H2SO4.8H2O} \SI{5}{\micro\m} & $1.7^{+2.5}_{-1.2}\%$ & $0.7^{+1.1}_{-0.5}\%$ & $0.8^{+1.3}_{-0.6}\%$ & $0.6^{+1.0}_{-0.4}\%$\\
	\ce{H2SO4.8H2O} \SI{50}{\micro\m} & $17.6^{+8.6}_{-9.0}\%$ & $23.3^{+5.7}_{-5.7}\%$ & $45.1^{+6.4}_{-6.6}\%$ & $43.2^{+4.7}_{-4.8}\%$\\
	\hline
	\ce{Mg(ClO3)2.6H2O} & $1.3^{+2.1}_{-1.0}\%$ & $0.5^{+1.0}_{-0.4}\%$ & $1.7^{+2.5}_{-1.3}\%$ & $2.6^{+3.9}_{-1.9}\%$\\
	\ce{Mg(ClO4)2.6H2O} & $1.5^{+2.2}_{-1.1}\%$ & $0.6^{+1.2}_{-0.4}\%$ & $1.8^{+3.0}_{-1.3}\%$ & $4.2^{+4.7}_{-2.9}\%$\\
	\ce{MgCl2.nH2O} & $5.6^{+3.5}_{-2.5}\%$ & $2.6^{+1.5}_{-1.2}\%$ & $6.5^{+4.8}_{-3.0}\%$ & $10.2^{+7.4}_{-4.6}\%$\\
	\ce{MgSO4} & $2.1^{+3.5}_{-1.6}\%$ & $1.1^{+1.8}_{-0.9}\%$ & $1.1^{+1.5}_{-0.8}\%$ & $0.9^{+1.4}_{-0.7}\%$\\
	\ce{MgSO4.nH2O} & $4.6^{+3.5}_{-2.4}\%$ & $2.7^{+2.1}_{-1.5}\%$ & $2.3^{+2.1}_{-1.3}\%$ & $2.3^{+2.0}_{-1.3}\%$\\
	\ce{Na2Mg(SO4)2.4H2O} & $10.2^{+6.0}_{-4.5}\%$ & $13.3^{+5.1}_{-5.1}\%$ & $4.7^{+3.1}_{-2.2}\%$ & $5.8^{+3.4}_{-2.3}\%$\\
	\ce{Na2SO4.10H2O} & $5.7^{+6.3}_{-4.0}\%$ & $18.4^{+3.9}_{-4.0}\%$ & $2.6^{+3.9}_{-2.0}\%$ & $2.3^{+2.8}_{-1.7}\%$\\
	\ce{NaClO4} & $0.8^{+1.5}_{-0.6}\%$ & $0.3^{+0.5}_{-0.3}\%$ & $0.7^{+1.1}_{-0.5}\%$ & $0.5^{+0.8}_{-0.4}\%$\\
	\ce{NaClO4.2H2O} & $1.3^{+1.9}_{-0.9}\%$ & $0.5^{+0.8}_{-0.4}\%$ & $1.5^{+2.4}_{-1.2}\%$ & $1.9^{+3.0}_{-1.4}\%$\\
	\ce{NaCl} & $1.0^{+1.6}_{-0.7}\%$ & $0.4^{+0.6}_{-0.3}\%$ & $0.7^{+1.3}_{-0.6}\%$ & $0.6^{+0.9}_{-0.4}\%$\\
	\hline
	All ices & $33.0^{+5.8}_{-5.3}\%$ & $30.4^{+3.6}_{-3.7}\%$ & $22.4^{+5.2}_{-5.5}\%$ & $17.5^{+4.2}_{-4.0}\%$\\
	All acids & $27.1^{+6.6}_{-7.5}\%$ & $27.1^{+5.1}_{-5.1}\%$ & $49.1^{+5.1}_{-6.0}\%$ & $46.1^{+3.9}_{-4.6}\%$\\
	All salts & $39.7^{+7.7}_{-6.7}\%$ & $42.2^{+5.5}_{-5.2}\%$ & $29.0^{+5.2}_{-5.0}\%$ & $36.6^{+4.2}_{-4.1}\%$\\
	\enddata
\end{deluxetable}

\bibliography{Europa Paper arXiv.bbl}{}
\bibliographystyle{aasjournal}
\end{document}